\documentclass[fleqn]{2023SCGE}
\newcommand\fs{\mbox{$.\!\!^{\mathrm s}$}}
\newcommand\farcs{\mbox{$.\!\!^{\prime\prime}$}}%

\def\swift{\emph{Swift}}
\def\flux{\mbox{erg\nonbreak{}cm$^{-2}$\nonbreak{}s$^{-1}$}}
\def\arc{\mbox{$^{\prime\prime}$}}

\def\src{EP240408a}
\def\nonbreak{~}

\begin{document}

\ensubject{subject}

\ArticleType{Article}
\SpecialTopic{SPECIAL TOPIC: }
\Year{2023}
\Month{January}
\Vol{66}
\No{1}
\DOI{??}
\ArtNo{000000}
\ReceiveDate{January 11, 2023}
\AcceptDate{April 6, 2023}

\title{Einstein Probe discovery of EP240408a: a peculiar X-ray transient with an intermediate timescale}{Einstein Probe discovery of EP240408a: a peculiar X-ray transient with an intermediate timescale}


\author[1]{Wenda Zhang}{{wdzhang@nao.cas.cn}}
\author[1,2]{Weimin Yuan}{{wmy@nao.cas.cn}}
\author[1,2]{Zhixing Ling}{{lingzhixing@bao.ac.cn}}
\author[3]{Yong Chen}{{ychen@ihep.ac.cn}}
\author[4,5]{Nanda Rea}{{rea@ice.csic.es}}
\author[6]{Arne Rau}{arau@mpe.mpg.de}
\author[7]{\\Zhiming Cai}{}
\author[1]{Huaqing Cheng}{}
\author[4,5]{Francesco Coti Zelati}{}
\author[8]{Lixin Dai}{}
% ...

\AuthorMark{W. Zhang, et al.}
\AuthorCitation{W. Zhang, W. Yuan, Z. Ling, Y. Chen, N. Rea, A. Rau, et al}

\address[1]{National Astronomical Observatories, Chinese Academy of Sciences, Beijing 100101, People’s Republic of China}
\address[2]{School of Astronomy and Space Sciences, University of Chinese Academy of Sciences, Beijing 100049, People’s Republic of China}
\address[3]{Key Laboratory of Particle Astrophysics, Institute of High Energy Physics, Chinese Academy of Sciences, Beijing 100049, People’s Republic of China}
\address[4]{Institute of Space Sciences (ICE), CSIC, Barcelona E-08193, Spain}
\address[5]{Institut d’Estudis Espacials de Catalunya (IEEC), Barcelona E-08034, Spain}
\address[6]{Max Planck Institute for Extraterrestrial Physics, Garching 85748, Germany}
\address[7]{Innovation Academy for Microsatellites, Chinese Academy of Sciences, Shanghai 201210, People’s Republic of China}
\address[8]{Department of Physics, University of Hong Kong, Hong Kong, People’s Republic of China}
% ...


\abstract{We report the discovery of a peculiar X-ray transient, EP240408a, by \textit{Einstein Probe (EP)} and follow-up studies made with \textit{EP}, \textit{Swift}, \textit{NICER}, GROND, ATCA and other ground-based multiwavelength telescopes.
The new transient was first detected with Wide-field X-ray Telescope (WXT) on board \textit{EP} on April 8th, 2024, manifested in an intense yet brief X-ray flare lasting for 12\nonbreak{}seconds.
The flare reached a peak flux of $3.9\times 10^{-9}~\rm erg~cm^{-2}~s^{-1}$ in 0.5--4\nonbreak{}keV,  $\sim$300 times brighter than the underlying X-ray emission detected throughout the observation. 
Rapid and more precise follow-up observations by \textit{EP}/FXT, \textit{Swift} and \textit{NICER} confirmed the finding of this new transient.
Its X-ray spectrum is non-thermal in 0.5--10\nonbreak{}keV, with a power-law photon index varying within 1.8--2.5. 
The X-ray light curve shows a plateau lasting for $\sim$4\nonbreak{}days, followed by a steep decay till becoming undetectable $\sim$10\nonbreak{}days after the initial detection.
Based on its temporal property and constraints from previous \textit{EP} observations, an unusual timescale in the range of 7--23\nonbreak{}days is found for \src, which is intermediate between the commonly found fast and long-term transients. 
No counterparts have been found in optical and near-infrared, with the earliest observation at 17 hours after the initial X-ray detection, suggestive of intrinsically weak emission in these bands.
We demonstrate that the remarkable properties of \src{} are inconsistent with any of the transient types known so far, by comparison with, in particular, jetted tidal disruption events, gamma-ray bursts, X-ray binaries and fast blue optical transients. The nature of \src{} thus remains an enigma.
We suggest that \src{} may represent a new type of transients with intermediate timescales of the order of $\sim$10 days. 
The detection and follow-ups of more of such objects are essential for revealing their origin.
}



\maketitle


\begin{multicols}{2}
\section{Introduction}\label{sec:intro}

The X-ray sky is rich in variable and transient objects. 
Since the early days of X-ray astronomy, X-ray transients of diverse types have been found, and their variable or transient activities occur on a wide range of timescales spanning many orders or magnitude. 
In the long-timescale regime, the most well-studied X-ray transients are transient X-ray binaries (XRBs; also called X-ray novae) fuelled by accretion onto stellar-mass black holes or neutron stars, which occasionally enter outbursts that typically last months to a few years (see, e.g., refs.~\cite{remillard_x-ray_2006}); during the outbursts they exhibit strong variability over a large range of timescales down to as short as milliseconds. In recent years, a few dozens of X-ray transients associated with (mostly) the nuclei of galaxies lasting for from months to years have been found, which are interpreted as tidal disruption events (TDEs) where a star is disrupted and partially accreted onto a massive black hole \cite{rees_tidal_1988,bade_detection_1996,komossa_discovery_1999,komossa_giant_1999,komossa_huge_2004}; in this process, strong gravity plays an important role in both the dynamics of the debris (essential for the early evolution of TDEs) and the X-ray emission (see refs.~\cite{komossa_tidal_2015, saxton_x-ray_2020} for a review).  

On the other end of the spectrum, fast X-ray transients or intense short bursts with durations from sub-seconds to hundreds of seconds, and to hours have been found in large numbers for several decades (e.g.~\cite{pye_ariel_1983,arefiev_fast_2003}). 
Though being of heterogeneous origins, most of these events are believed to be produced in extremely violent processes on or involving stellar-mass objects.
The most common examples are the X-ray counterparts or afterglows of gamma-ray bursts (GRBs) produced by collapse of massive stars or merger of two neutron stars~\cite{zhang_physics_2018}, as well as X-ray flashes~\cite{heise_x-ray_2001} that are thought to be the variants of GRBs. 
Others include flares from magnetars with ultrastrong magnetic fields~\cite{rea_magnetar_2011,turolla_magnetars:_2015}, 
supernova shock breakouts~\cite{soderberg_extremely_2008,alp_blasts_2020}, X-ray flares arising from coronal activities of stars lasting for mostly from within an hour to a day~\cite{benz_physical_2010}. 
In recent years, samples of fast X-ray transients at fainter X-ray fluxes have also been found in deep X-ray surveys, some of unknown nature albeit with possible models suggested (\cite{bauer_new_2017,xue_magnetar-powered_2019}), fast X-ray transients found in \textit{Chandra} and \textit{XMM-Newton} surveys, e.g.~\cite{quirola-vasquez_extragalactic_2022,quirola-vasquez_extragalactic_2023}.

In contrast, somewhere in between fast and long-term transients, X-ray transients with intermediate timescales appear to be comparably rarer, however.
One interesting example in this timescale range is CXOU J005245.0-722844, a rare BeWD binary in the Small Magellanic Cloud, which entered a recent outburst lasting for only 6--12\nonbreak{}days~\cite{marino_einstein_2024}.

The \textit{Einstein Probe (EP)}~\cite{yuan_einstein_2022}\footnote{Einstein Probe is a international collaborative mission led by the Chinese Academy of Sciences and participated by ESA, MPE and CNES.} is a space X-ray observatory dedicated to the discovery and characterization of X-ray transients.
One of the two payloads onboard is the Wide-field X-ray Telescope (WXT), a soft X-ray focusing telescope with a $\sim$3800\nonbreak{}sqr.~deg.\ field-of-view (FoV) enabled by novel technology of lobster-eye micro-pore optics. 
Operating in the 0.5-4\,keV band, WXT features a significantly improved sensitivity ($(2-3) \times 10^{-11}~\rm erg~cm^{-2}~s^{-1}$ at 1\,ks exposure) over the current wide-field X-ray monitors, making it a powerful instrument in detecting relatively faint X-ray transients in soft X-rays.
Another instrument is the Follow-up X-ray Telescope (FXT) \cite{chen_status_2020} consisting of two co-aligned identical units using X-ray mirrors built from the Wolter-I optics and pn-CCDs as focal-plane detectors. 
FXT has a FoV of $1 \times 1$ square degrees, a spatial resolution of $20-24$\,arcsec (half-power diameter, on-axis) and an effective area of $\sim 300$\,cm$^2$ at 1\,keV (one unit).
\textit{EP} was launched on January 9th, 2024, followed by a commissioning and calibration phase of about 6 months. 
\textit{EP} has started scientific operations since July, 2024.

On April 8th, 2024, a new X-ray transient, later dubbed EP240408a, was detected by \textit{EP}/WXT in a calibration observation for FXT during the commissioning phase \cite{hu_ep240408a_2024}. 
A series of follow-up observations with space- and ground-based telescopes at multi-wavelengths were triggered. 
In this work, we present the discovery, results of follow-up observations and discussion about the nature of this transient. The paper is organized as follows. In Section~\ref{sec:data} we present the observations and the data reduction. The results are presented in Section~\ref{sec:results}. Based on the results, we discuss possible scenarios for the source in Section~\ref{sec:discussion} and summarise our work in Section~\ref{sec:conclusion}.



\section{Observations and data reduction}
\label{sect:data}

On April 8th, 2024, an uncatalogued new X-ray source, designated \src{}, was detected by \textit{EP}/WXT during an observation in the commissioning phase~\cite{hu_ep240408a_2024}. We triggered prompt follow-up observations with the \textit{Neil Gehrels Swift Observatory (Swift)} \cite{gehrels_swift_2004} and \textit{EP}/FXT, $\sim$33 and $\sim$42\nonbreak{}hours after the WXT detection, respectively.
The source was further detected with much improved localization and spectral quality~\cite{hu_ep240408a_2024-1,rea_ep240408a_2024} by both, the X-ray Telescope (XRT) \cite{burrows_swift_2005} on board  \textit{Swift} and FXT; whilst no counterpart was detected in the optical to near-ultraviolet (NUV) band by the Ultra-Violet/Optical Telescope (UVOT) \cite{roming_swift_2005}.  
\src{} was also observed with the \textit{Neutron star Interior Composition Explorer (NICER)} \cite{gendreau_neutron_2016} for several times, with the first observation carried out $\sim$1.7 days after the WXT detection.
The journal of the X-ray observations is given in Table~\ref{tab:log}. The source position was later observed with ground-based multi-wavelength telescopes in the optical, near-infrared (NIR) and radio bands, but no counterpart was detected, however. 
The observations and data reduction and analysis are presented in details in below.

\label{sec:data}
\subsection{\textit{Einstein Probe X-ray observations}}
\label{sec:ep}


\subsubsection{\textit{\textit{EP}/WXT}}

\begin{figure*}
    \includegraphics[width=\textwidth]{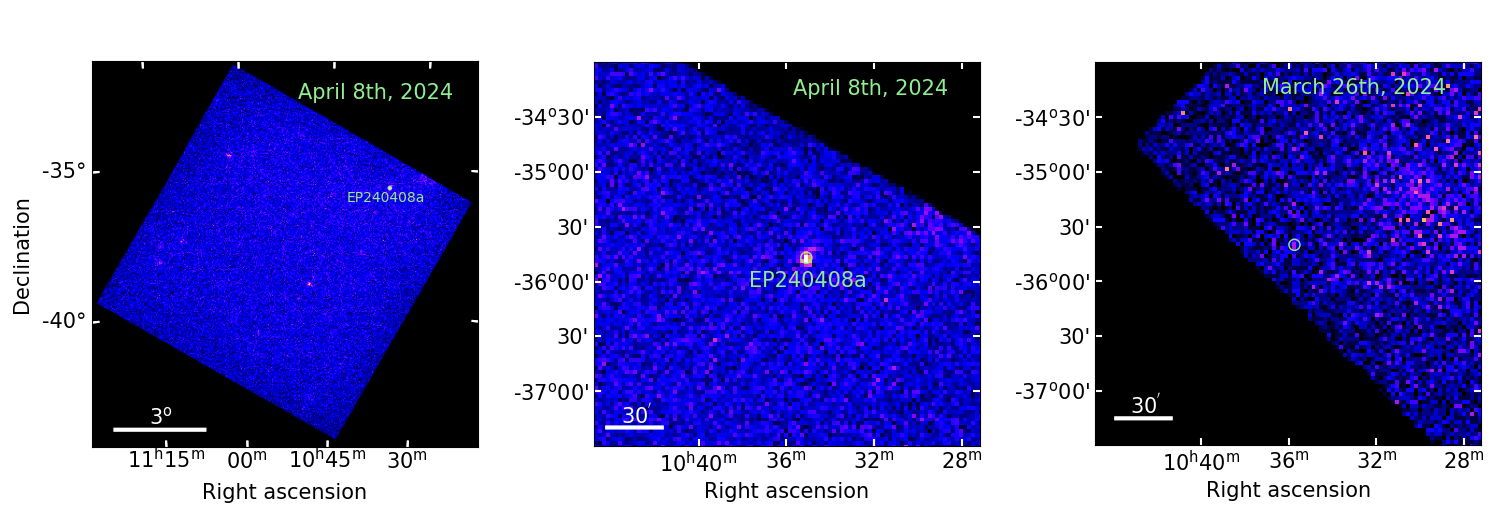}
    \caption{Discovery X-ray images of \src{} in 0.5--4\nonbreak{}keV obtained with \textit{EP}/WXT: on the whole CMOS detector chip (left) and a close-up (middle). An X-ray image at the source position extracted from an observation taken $\sim$13\nonbreak{}days before the discovery is also shown for comparison (right). In all panels, the WXT error circle of 3\,arcmin in radius is marked by a green circle at the position of \src.\label{fig:wxtimg}}
\end{figure*}

\src{} was first detected on the CMOS sensor No.~29 of WXT in an observation (ObsID 13600005155) starting at April 8th, 2024 at 12:14:12 UT and lasting for $83.8~\rm ks$. The image is shown in the middle panel of Fig.~\ref{fig:wxtimg}, in which the source is clearly seen. The observation was divided into 20 segments due to Earth occultation and passage through the South Atlantic Anomaly (SAA), resulting in a net exposure of $38.5~\rm ks$. The WXT data are processed with \texttt{wxtpipeline}, an analysis chain of the WXT Data Analysis Software (\texttt{WXTDAS}) developed at the EP Science Center (EPSC). 
The \texttt{wxtpipeline} tool calibrates the raw data by performing coordinate transformation, flagging hot and bad pixels, computing the pulse invariant values, filtering the events to obtain a clean event list using the default criteria, and extracting a sky image from the clean event list. For the analysis, the events with grade patterns of 0--12 are selected.
Then the tool carries out detection of point-like sources and extraction of the various products of detected sources, including the light curves and spectra, the ancillary response file and response matrix, as well as the background light curves and spectra. 
By default, the light curves and spectra of the sources are extracted from a circular region with a radius of $9.1^\prime$, and those of the background are extracted from an annulus with inner and outer radii of $18.2^\prime$ and $36.4^\prime$, respectively.

In this observation, \src{} was detected with a net count of 255.3 at a significance of $15.3\sigma$ in 0.5--4\nonbreak{}keV. The WXT position of the source is R.A. = $10^{\mathrm h}35^{\mathrm m}23.9^{\mathrm s}$, Decl. = $-35^\circ44^{\prime}55{\farcs}2$ (J2000) with an uncertainty of $3^\prime$ at the $90\%$ confidence level (c.l.). There is no previously known X-ray source at this position, suggesting \src{} to be a new X-ray transient. Thanks to the large field-of-view of WXT, several follow-up pointings covering the sky position of \src{} were conducted, with exposure times ranging from 10.1 to $26.3~\rm ks$ (see Table~\ref{tab:log}). The source was significantly detected till the observation taken on April 11th, 2024 (ObsID 08500000059).
Since then \src{} has remained no longer detected with WXT.

Since \src{} was already present from the start of the discovery observation (ObsID 13600005155), it may have started to brighten before the observation. 
A search of the \textit{EP}/WXT data at earlier epochs shows that the source position has been covered by WXT several times since March, 2024, at which no X-ray emission was detected, however.  
The latest observation (ObsID 08503145728) prior to the first detection was taken on March 26th, 2024 with an exposure time of $10.0~\rm ks$, imposing an upper limit of 2.6$\times 10^{-3}~\rm  counts~ s^{-1}$ (90\% c.l.). The weakness of \src{} is obvious by looking at the image extracted from this observation, as shown in the right panel of Fig.~\ref{fig:wxtimg}.
Therefore, the onset of the transient is likely to occur sometime between March 26th and April 8th. Hereafter, we refer to the start of the first discovery observation, 2024-04-08UT12:14:12, as $t_0$, the trigger time of \src{}.
 


\subsubsection{\textit{\textit{EP}/FXT}}

The \textit{EP}/FXT consists of two identical and co-aligned telescope units: FXT-A and FXT-B.
FXT observed \src{} twice (Table~\ref{tab:log}), with both FXT-A and FXB-B operating in the full-frame (FF) mode that has a time resolution of $\simeq$50\nonbreak{}ms. 
The first observation, beginning $\sim$1.8\nonbreak{}days (ObsID: 08500000058) after the initial detection by WXT, has an elapsed time of $\sim$49\nonbreak{}ks and an on-source time of $\sim$24\nonbreak{}ks, while the second observation (ObsID: 08500000116) was performed much later, on June 25th, 2024, with an exposure of $7.7~\rm ks$.
The data are processed by using the \texttt{fxtchain} tool, as part of the FXT Data Analysis Software (\texttt{FXTDAS}) developed at the EPSC. This tool performs several tasks, including identifying particle events, calculating pulse-invariant values for the event files, flagging bad and hot pixels in the FXT detectors, and selecting good time intervals using housekeeping data. It then extracts the spectra and light curves for both the source and background, and produces ancillary files and redistribution matrices.

\begin{figure*}
 \includegraphics[width=\textwidth]{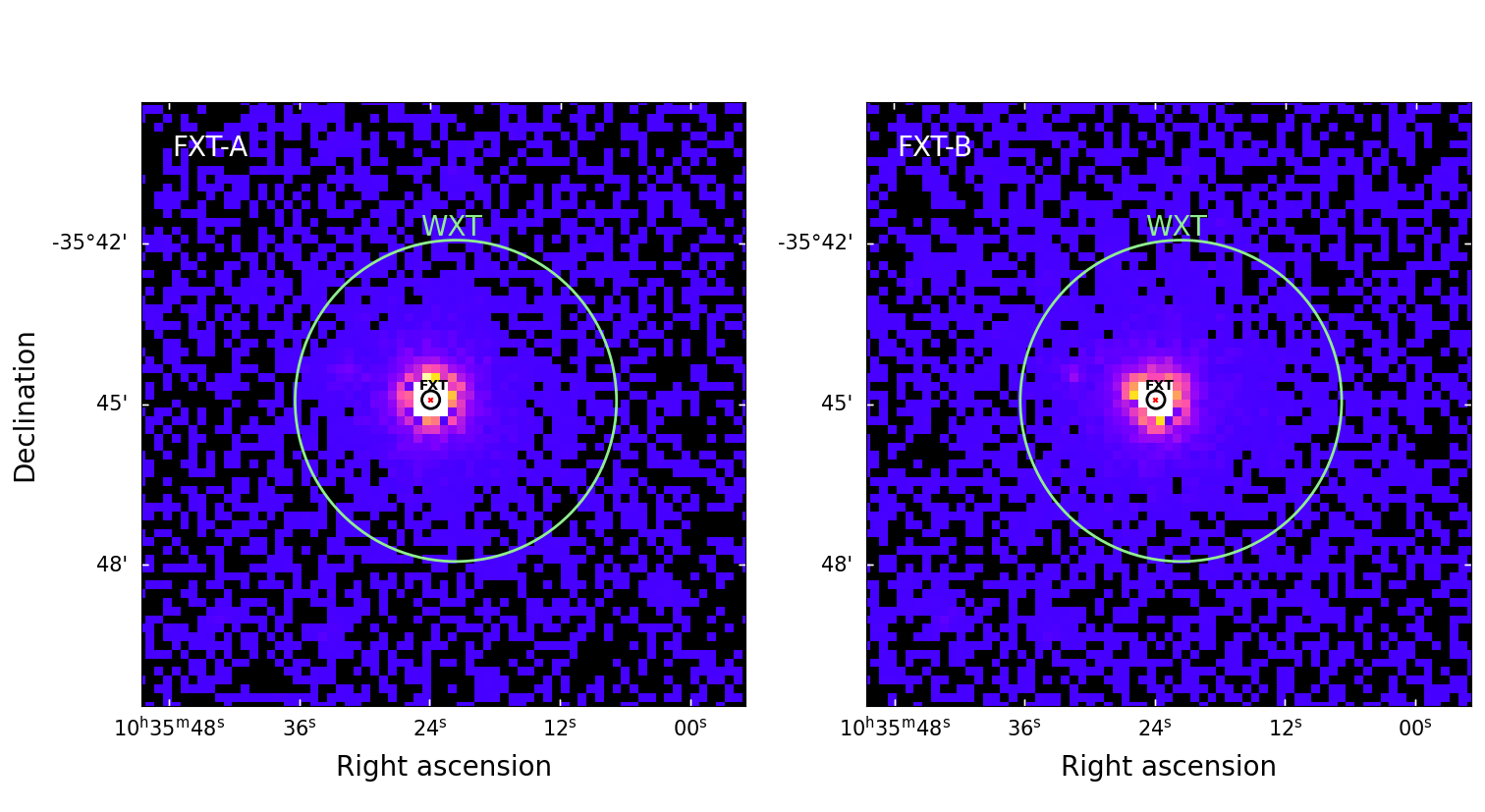}
 \caption{X-ray images of \src{} obtained with the two \textit{EP}-FXT units, FXT-A (left) and FXT-B (right), extracted from the observation taken $\sim$42\nonbreak{}hours after the discovery by WXT. The WXT and FXT localizations are indicated by green and black circles, respectively, while the XRT position with an error radius of $2.2"$ is indicated by a red cross.\label{fig:fxt_img}}
\end{figure*}

In the first FXT observation, an X-ray source was detected within the WXT positional error circle of $3^\prime$ ($90\%$ c.l.), suggestive of being consistent with the WXT source. 
The FXT observation also improves significantly the source localization, reaching a precision of 5\nonbreak{}arcsec (90\% c.l.).
The images of the FXT detection are shown in Fig.~\ref{fig:fxt_img}.
The net source count is $2.4\times 10^4$ for FXT-A and $2.7\times 10^4$ for FXT-B, giving a source count rate of $0.94~\rm counts~s^{-1}$ and $1.05~\rm counts~s^{-1}$, respectively.
The source was not detected in the second FXT observation taken on June 25th, 2024, setting a $3\sigma$ upper limit on the source count rate as $1.84\times 10^{-3}~\rm counts~s^{-1}$ in 0.5--10\nonbreak{}keV.

\begin{table*}
\caption{
\label{tab:log}
Journal of the X-ray observations of \src\ presented in this work.}
\centering
\begin{tabular}{lccc}
\hline\hline
Telescope/Instrument (Mode)	& Obs ID 		&Start -- End time	        		   & Exposure  \\
					   &			& Mmm DD hh:mm:ss (UTC) & (ks)    \\
\hline
EP/WXT-29& 13600005155 & Apr 8  12:14:12 -- Apr 8 11:44:06     & 38.5  \\
Swift/XRT (PC)	 & 00016599001		    & Apr 10 02:57:40 -- Apr 10 04:48:35    & 1.8	\\
EP/WXT-14  & 08500000058    & Apr 10 12:30:12 -- Apr 11 04:00:54 & 26.2 \\
EP/WXT-37  & 08500000058    & Apr 10 12:30:12 -- Apr 11 04:00:54 & 26.3 \\
EP/FXT (FF)		 & 08500000058 		    & Apr 10 12:36:18 -- Apr 11 02:17:06    & 24.0  \\
NICER/XTI        & 7204340101			& Apr 10 18:30:31 -- Apr 10 21:40:18 & 0.3   \\
NICER/XTI        & 7204340102			& Apr 11 02:08:20 -- Apr 11 02:18:54 & 0.5   \\
EP/WXT-13 & 08500000059 & Apr 11 20:35:44 -- Apr 12 08:06:03 & 14.4 \\
EP/WXT-38 & 08500000059 & Apr 11 20:35:44 -- Apr 12 08:06:03 & 10.1 \\
NICER/XTI        & 7204340103			& Apr 12 09:07:29 -- Apr 12 18:51:20   	& 2.0 \\
NICER/XTI        & 7204340104			& Apr 13 00:49:45 -- Apr 13 20:51:00  	& 1.4 \\
NICER/XTI        & 7204340105			& Apr 14 04:29:47 -- Apr 14 12:16:01   	& 0.3  	\\
NICER/XTI        & 7204340106			& Apr 15 08:22:22  -- Apr 15 08:29:54	& 0.5 	\\
Swift/XRT (PC)	 & 00016599003		    & Apr 19 03:41:27 -- Apr 20 03:27:53   	& 1.0	\\
Swift/XRT (PC)	 & 00016599004		    & Apr 21 03:03:55 -- Apr 21 04:50:52    & 1.2 	\\
NICER/XTI        & 7204340113			& Apr 22 01:22:59 -- Apr 22  01:27:45   & 0.3 	\\
Swift/XRT (PC)	 & 00016599005		    & Apr 24 03:53:46 -- Apr 24 10:04:52    & 1.0 	\\
Swift/XRT (PC)	 & 00016599006		    & Apr 26 04:40:59 -- Apr 27 07:31:52   	& 2.4 	 \\
EP/FXT (FF) & 08500000116         & Jun 25 07:07:09 -- Jun 25 10:54:29    & 7.7 \\
\hline
\hline
\end{tabular}
\end{table*}

\subsection{\textit{Swift observations in X-ray, optical and UV}}
\label{sec:swift}
The \textit{Swift} observatory observed \src\ five times from April 10th to April 27th (PI: Jingwei Hu; see Table~\ref{tab:log}), with the first observation taken  $\sim33$ hours after the initial WXT detection.
For all the observations, the \textit{Swift}/XRT was configured in the photon counting (PC) mode. 
During the first observation on April 10th, an X-ray source was detected at a position consistent with that of the FXT source (thus within the WXT error circle, see Fig.~\ref{fig:fxt_img}), suggesting its association with \src{}. 
The XRT observation gives a further improved source position, enhanced by applying correction using the \textit{Swift}/UVOT astrometry \cite{evans_methods_2009}: R.A. = 10$^\mathrm{h}$35$^\mathrm{m}$23$\fs$96, Decl. = $-$35$^{\circ}$44$^{\prime}$55$\farcs$1 (J2000; uncertainty of 2.2\arc\ at the 90\% c.l.)\footnote{In the GCN reporting the XRT observation \cite{hu_ep240408a_2024-1}, the XRT localization was reported without correction using the UVOT astrometry. The new XRT position is now $5.5\arc$ away from the position that was reported in that GCN.}. 
The FXT source position is well consistent with the XRT position (marked as a red cross in Fig.~\ref{fig:fxt_img}), with a separation of only 0.7\nonbreak{}arcsec, though its uncertainty of $\sim 5\arc$ is slightly larger.
We thus adopt the XRT coordinates as the X-ray position of \src{}. 

The light curves and spectra are extracted from a circular aperture with a radius of 47.2 arcsec for the source, and an annular region with inner and outer radii of 94.4 and 188.8 arcsec, respectively, for the background. 
In this observation, the source has a total of 654.5 net source counts and an averaged count rate of 0.36$\pm0.01~\rm counts~s^{-1}$ in 0.5--10\nonbreak{}keV.  
In the following four observations, \src{} was no longer detected by XRT.  
We derive 3$\sigma$ upper limits on the net source count rate ranging from 0.005 to $0.01~\rm counts~s^{-1}$ (0.5--10\nonbreak{}keV), depending on the duration of the observation. By combining the data from these observations, a more stringent upper limit of $0.003~\rm counts~s^{-1}$ (0.5--10\nonbreak{}keV) are set.

During the first observation, the \textit{Swift}/UVOT operated in the image mode and utilized all six of its filters. \src\ was not detected in any of these filters, however, down to the following magnitude limits (the Vega system): UVW2 $>$ 19.5\nonbreak{}mag, UVM2 $>$ 18.9\nonbreak{}mag, UVW1 $>$ 18.6\nonbreak{}mag, U $>$ 19.4\nonbreak{}mag, B $>$ 19.8\nonbreak{}mag, V $>$ 18.9\nonbreak{}mag.

\subsection{\textit{NICER X-ray observations}}
\label{sec:nicer}
The X-ray Timing Instrument (XTI) onboard \textit{NICER} observed \src\ 32 times  (PIs: Nanda Rea, Brendan O'Connor), first beginning on April 10th at 18:30:31 UT, $\sim$41 hours after the initial WXT detection.
The data are processed using \texttt{nicerl2} with specific options $overonly\_range=0-5$ and $cor\_range="1.5-*"$, to minimize potential contamination from non-X-ray flares\footnote{\url{https://heasarc.gsfc.nasa.gov/docs/nicer/analysis\_threads/flares/}}. After applying such a strict selection criterion, only 6 out of the first 12 observations  (7204340101-7204340112) are left with clean events. 
From the observation on April 22nd (ObsID 7204340113) onward, the source was no longer detected with XTI. For brevity, only the NICER observations where \src{} is detected are listed in Table~\ref{tab:log}.

The photon arrival times are corrected to the barycenter of the Solar system using the XRT position and the JPL DE430 ephemeris. Light curves and spectra are extracted using the \texttt{nicerl3-lc} and \texttt{nicerl3-spec} pipelines, respectively, employing the \texttt{SCORPEON} background model\footnote{\url{https://heasarc.gsfc.nasa.gov/docs/nicer/analysis_threads/scorpeon-overview/}}.
For spectral fitting, we fit the source and background spectra jointly in the energy range of 0.5--12\nonbreak{}keV, with the background model parameters $niscorpv23\_noisenorm$, $niscorpv23\_skylhb\_em$ and $niscorpv23\_skygal\_nh$ fixed. The background parameter $niscorpv23\_swcxok\_norm$ is left free to vary if an excess towards low energies is seen around $\sim 0.6~\rm keV$, to model the oxygen K$\alpha$ emission by the solar wind.


\subsection{Upper limits from other high-energy missions}
\label{sec:erosita}
eROSITA: 
The position of \src\ was covered by eROSITA \cite{predehl_erosita_2021} on board the Spektrum Roentgen Gamma observatory for $\sim$127\nonbreak{}s during the first six months of survey operations \cite{merloni_srgerosita_2024}. No X-ray emission was detected at the source position, with a flux upper limit of 6$\times$10$^{-14}$\nonbreak{}\flux\ (0.2--2.3\nonbreak{}keV; see refs.~\cite{tubin-arenas_erosita_2024}).


GECAM: The Gravitational wave high-energy Electromagnetic Counterpart All-sky Monitor (GECAM) is a GRB monitor operating in the hard X-ray to gammay-ray band,  consisting of two identical micro-satellites in opposite orbital phases \cite{li_gecam_2020}.
One of the satellite modules, GECAM-B, covered the sky position of \src{} at the time of the first WXT detection~\cite{wang_ep240408a_2024}. No sources were detected and upper limits of $\sim$(2--4)$\times 10^{-6}~\rm erg~cm^{-2}$ were imposed on the 15--300\nonbreak{}keV fluence, varying with the spectral model assumed. The fluence is integrated with respective to the time interval of the soft X-ray flare detected in the WXT light curve of \src{} during the first-detection observation (see below).

\subsection{Optical and near-infrared observations}
\label{sec:grond}

\begin{figure*}
\centering
\includegraphics[scale=0.23]{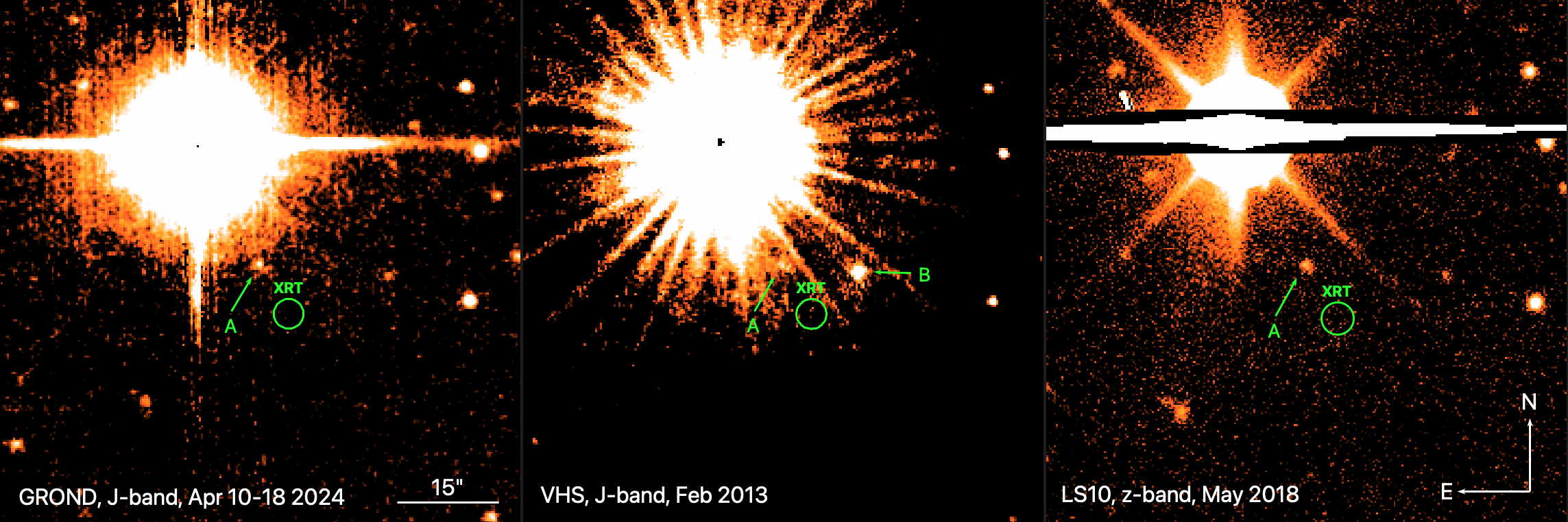}
\caption{{\it Left:} GROND J-band stack of the best-seeing data from April 10th, 12th, and 18th, 2024. A faint source (marked A) is detected 8.0\arc\ from the centre of the 2.2\arc\ radius \swift/XRT error circle (the green circle). {\it Middle:} Archival J-band image from the VISTA Hemisphere Survey (VHS) obtained in Feb 2013 showing emission consistent with source A. {\it Right:} Archival Legacy Survey DR10 (LS10) z-band image taken in May 2018 also showing source A. Note, there is a bright source, B, $\sim$9.3\arc\ to the West of the XRT position seen in the VHS data. No source is detected at this position by GROND and LS10.}
\label{Fig:grond_finder}
\end{figure*}

Follow-up observations were carried out with the \textit{Gamma-Ray Burst Optical Near-ir Detector (GROND)}~\cite{greiner_grond7-channel_2008} instrument mounted on the MPG 2.2m telescope at ESO's La Silla Observatory on April 10th, 12th, 17th and May 27th, 2024. These observations were performed simultaneously in the J, H, and K$_{\rm s}$-bands with typical exposures of $30~\rm min$ per epoch and band. The data are reduced using the standard IRAF-based GROND pipeline~\cite{kruhler_2175_2008}. The aperture photometry is calibrated against the Two Micron All Sky Survey catalogue (2MASS)~\cite{skrutskie_two_2006} and converted into the AB magnitude system.
No NIR source was found in the GROND images within the \swift/XRT error circle down to 3$\sigma$ limiting AB magnitudes of $J>21.4$, $H>21.2$, and K$_{\rm s}>19.6$. An unresolved source is detected in the J and H-bands at R.A.=10$^\mathrm{h}$35$^\mathrm{m}$24$\fs$31, Decl. = $-$35$^\circ$44$^{\prime}$47$\farcs$8 (J2000) with an uncertainty of 0.16\arc\ and 0.13\arc, respectively (source A in Fig.~\ref{Fig:grond_finder}). This source was initially suggested to be the counterpart to \src\ by refs.~\cite{rau_ep240410a_2024}, but it is now located 8.0\arc\ from the centre of the enhanced 2.2\arc\ radius XRT error circle.

In addition, several other optical observations were conducted, but none detected an optical counterpart. 
Observations made with GSP, MASTER and BOOTES-6/DPRT yielded upper limits of 
$21.5$\nonbreak{}mag in the r-band \cite{li_ep240408a_2024}, 19.5\nonbreak{}mag (unfiltered) \cite{antipov_ep240408a_2024}, and 20.5\nonbreak{}mag (clear filter) \cite{perez-garcia_ep240408a_2024}, taken $\sim$17\nonbreak{}hours, $\sim$23\nonbreak{}hours, and $\sim$25\nonbreak{}hours after the initial X-ray detection, respectively. The 2.56 m Nordic Optical Telescope (NOT; Roque de los Muchachos observatory, La Palma, Spain) performed late time observations on April 17th, 18th, and May 2nd (9.2, 10.2, and 24.1\nonbreak{}days after the initial X-ray detection), obtaining upper limits of 20.8, 21.4, and 22.0\nonbreak{}mag, respectively, in the z-band.


\subsection{ATCA radio observations}
\label{sec:atca}

We observed \src{} with the \textit{Australia Telescope Compact Array (ATCA)} on May 8th and 10th, 2024, $\sim$30 and $\sim$32\nonbreak{}days after the initial X-ray detection respectively, under the Director's Discretionary Time program with code CX570 (PI: Xinwen Shu).
The observations were first performed at the C-band and then at the K-band, using dual receivers with central frequencies of 5.5/9, and 16.7/21.2 GHz, respectively. All observations were taken with the ATCA CABB in the full 2048 spectral channel mode, resulting in 2 GHz of bandwidth.
To solve for the time-dependent complex gains, we used the nearby phase calibration source 1034-374 for C-band and 1048-313 for K-band, while the standard calibrators 0823-500 and 1034-293 were used as bandpass calibrators and to set flux density scale for C-band and K-band, respectively.

The data are reduced using the Common Astronomy Software Applications (CASA, version 5.6.1)\cite{mcmullin_casa_2007}. By loading the data using the {\tt IMPORTATCA} task, we examine each spectral window and flagged abnormal data due to the radio frequency interference or hardware issues. After checking and flagging the data, we set the flux scale for the bandpass calibrator using the {\tt setjy} task. Finally, we calibrate the data and then apply the solutions to \src{}. The calibrated data are then selected and we employ the {\tt CLEAN} algorithm to remove possible contamination from side-lobes, utilizing the conventional Briggs weighting with a ROBUST parameter of 0.5. The final cleaned maps exhibit an rms noise of $\sim$25\nonbreak{}$\mu$Jy/beam in the C-band and $\sim$40\nonbreak{}$\mu$Jy/beam in the K-band, measured using the {\tt IMFIT} task in CASA. We do not detect \src~in either C-band or K-band, resulting in 5$\sigma$ upper limits of $\sim$125\nonbreak{}$\mu$Jy and $\sim$200\nonbreak{}$\mu$Jy on the flux density, respectively.

\section{Observed properties of \src{}}
\label{sec:results}
\src{} is positioned at R.A. = 10$^\mathrm{h}$35$^\mathrm{m}$23$\fs$96, Dec. = $-$35$^{\circ}$44$^{\prime}$55$\farcs$1 (J2000; uncertainty of 2.2\arc\ at 90\% c.l.), determined by \textit{Swift}/XRT and consistent with the \textit{EP}/FXT measurement. 
No counterparts were detected in the other wavebands observed, however.
This section presents the X-ray temporal and spectral properties of \src{}, ending with a summary of the observed properties.

\subsection{X-ray temporal property}
\label{sec:detection}

\subsubsection{Short-term X-ray flare}

\begin{figure*}
 \includegraphics[width=\textwidth]{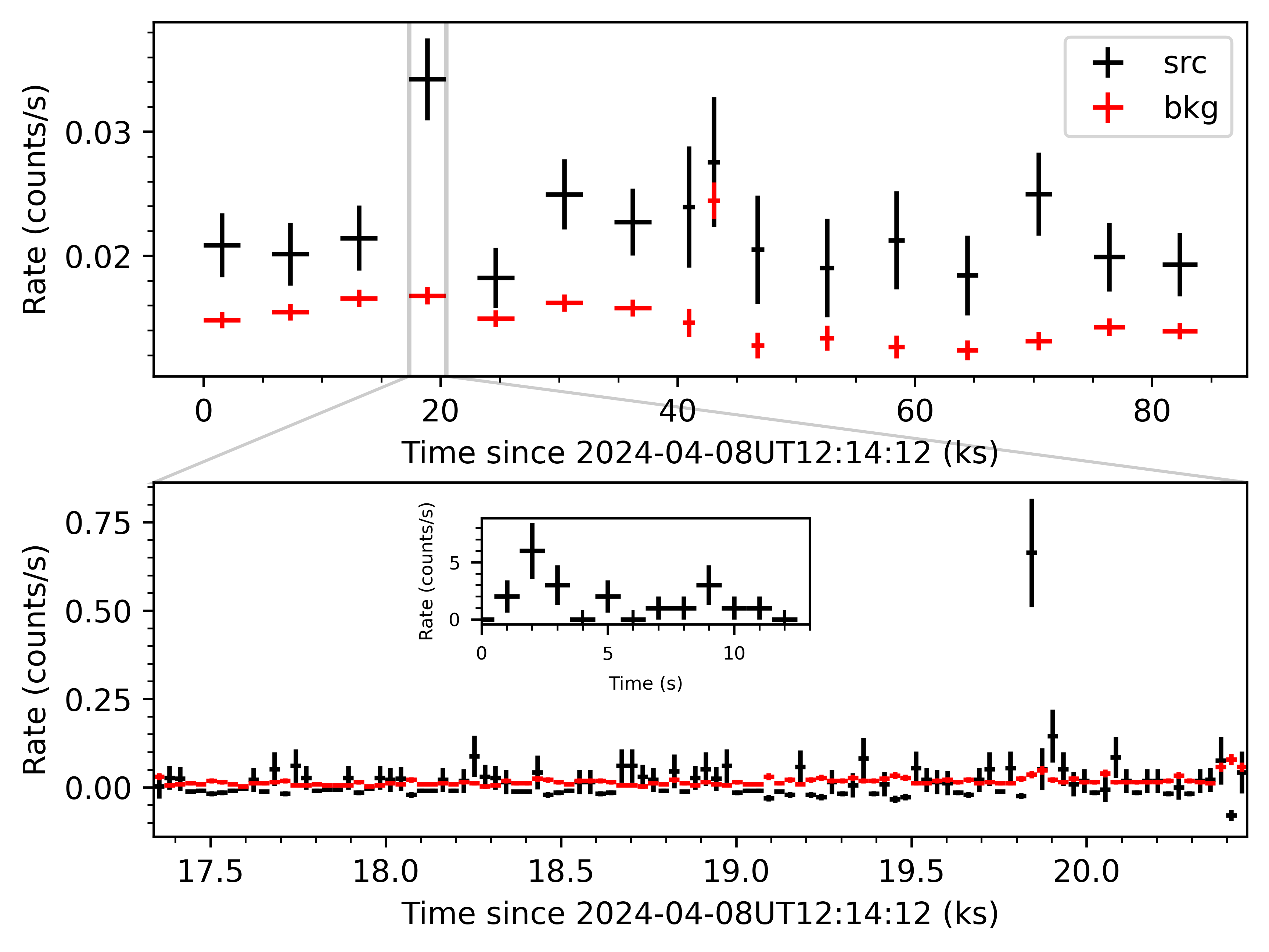}
 \caption{Light curves of \src{} of the WXT observation (13600005155) in which the transient was initially detected. Upper panel: the light curve of the entire observation. The background light curve (red) is also plotted for comparison. The time bins correspond to the orbital segments of the observation. For a few segments with a short duration, their light curves are not shown for clarity. Lower panel: a detailed light curve of the 4th segment, in which the short-term flare is located. The temporal bin size is $30~\rm s$. The inset shows the detailed light curve of the flare with a temporal bin size of $1~\rm s$.
 \label{fig:lc_gti}}
\end{figure*}

Fig.~\ref{fig:lc_gti} shows the X-ray light curve of \src{} in 0.5--4\nonbreak{}keV obtained during the \textit{EP}/WXT observation (ObsID 13600005155) in which the transient was first detected.
The overall light curve (upper panel in Fig.~\ref{fig:lc_gti}) spans over the length of the observation of 83.8\nonbreak{}ks, with time bins corresponding to the (orbital) segments.
Of particular interest, a short-term yet intense flare with a duration of $\sim$10\nonbreak{}s occurred clearly at an epoch $\sim$19.8\nonbreak{}ks from the start of the observation, as shown in a close-up light curve of the 4th segment with a bin size of 30\nonbreak{}s  (lower panel in Fig.~\ref{fig:lc_gti}). 
To study the flare in detail, we extract photon event lists of the source and background from the time segment in which the flare is located, and apply the Bayesian Block (BB)~\cite{scargle_studies_2013} algorithm to the event lists. 
As the background is also variable, in practice we utilize a modified version of the BB algorithm introduced by refs.~\cite{de_luca_extras_2021}, which accounts for the variation of the background.
The start time and the duration of the flare, as determined by the BB algorithm are 2024-04-08UT17:44:54.9 and $12.3~\rm s$, respectively. 
The average net count rate during the flare is $1.57$ counts\nonbreak{}s$^{-1}$, which is over two orders of magnitude higher than the persistent count rate of $6.78\times 10^{-3}$ counts\nonbreak{}s$^{-1}$.
A light curve of the flare with an even finer bin size of $1~\rm s$ is extracted and shown in the inset of the lower panel of Fig.~\ref{fig:lc_gti}. 
It shows that the flare possibly exhibits three peaks, each having a largely symmetric profile.

We find that the source was already significantly detected above the background even \textit{before} the short-term flare.
In fact, apart from the flare, the source did not show significant variation in 0.5--4\nonbreak{}keV during the overall observation of 83.8\nonbreak{}ks.
No similar flares are found in the light curves of the other X-ray observations of \src{}.

\subsubsection{Overall X-ray light curve}

\src{} was detected by \textit{EP}/WXT and \textit{NICER}/XTI multiple times, 
with WXT covering the early part and XTI covering the latter part of the observations. 
We demonstrate the variation of its X-ray emission in a model-independent approach by using the count rates and spectral hardness measured by each of the two instruments, respectively. 
Light curves are extracted in two energy bands, a soft (0.5--1\nonbreak{}keV) and a hard band (1--4\nonbreak{}keV for \textit{EP}/WXT and 1--5\nonbreak{}keV for \textit{NICER}/XTI). 
The choice of the dividing energy at $1~\rm keV$ is somewhat operational, so as to give roughly comparable numbers of counts in the two bands. 
We define hardness ratio as $(H-S)/(H+S)$, where $S$ and $H$ denote the \textit{background-subtracted} count rates in the soft and hard bands, respectively.

\begin{figure*}
    \centering
    \includegraphics[width=\textwidth]{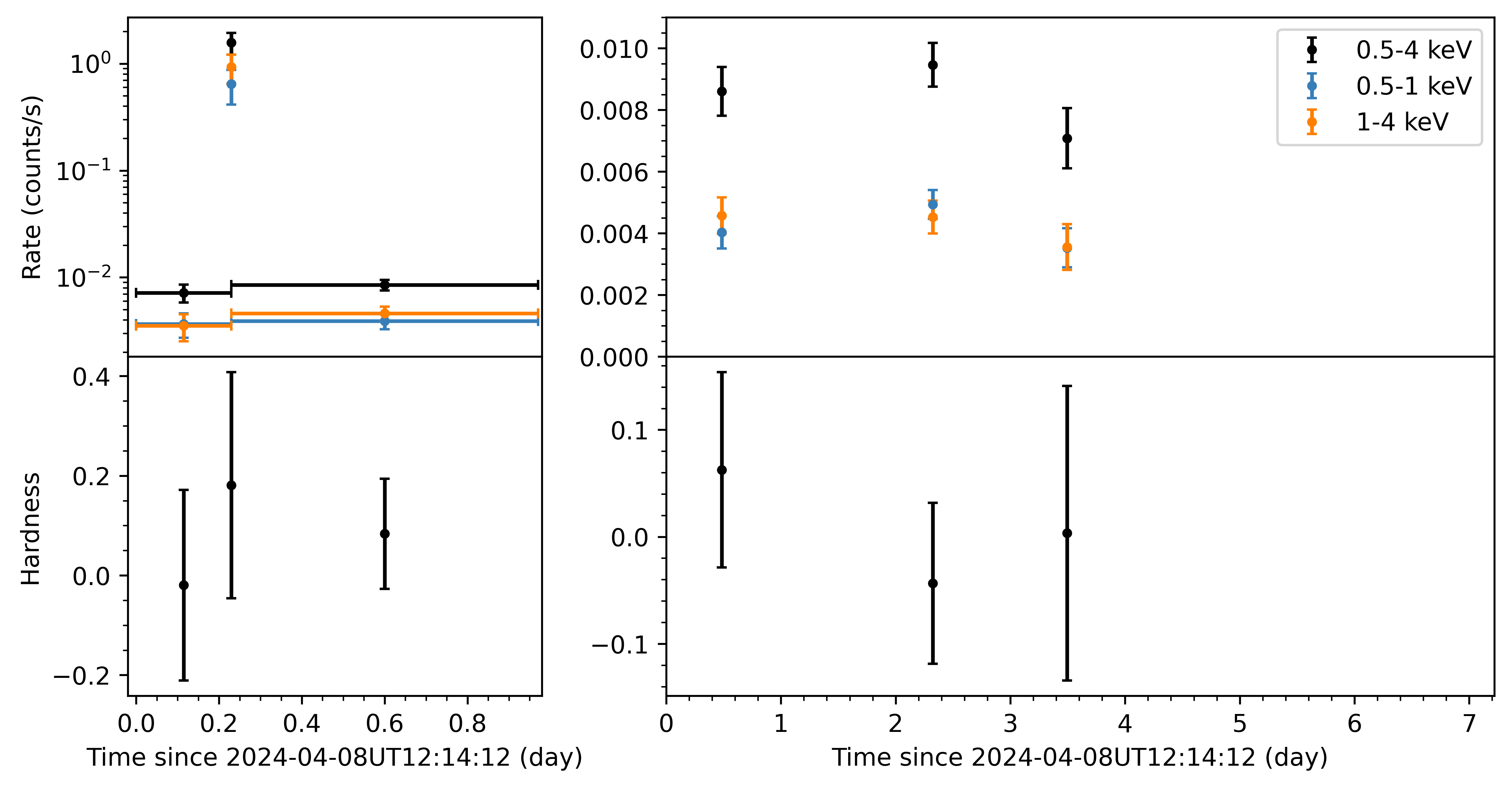}
    \caption{Left: the averaged count rates (upper) and hardness ratios (lower) in the pre-flare, flare and post-flare epochs during the first WXT observation. Right: averaged count rates (upper) and hardness ratio (lower) for the 3 WXT observations. Light curves of the soft (0.5--1\nonbreak{}keV; orange) and hard (1--4\nonbreak{}keV; blue) bands are shown. \label{fig:lc_wxt}}
\end{figure*}

\begin{figure*}
\centering
\includegraphics[width=.8\textwidth]{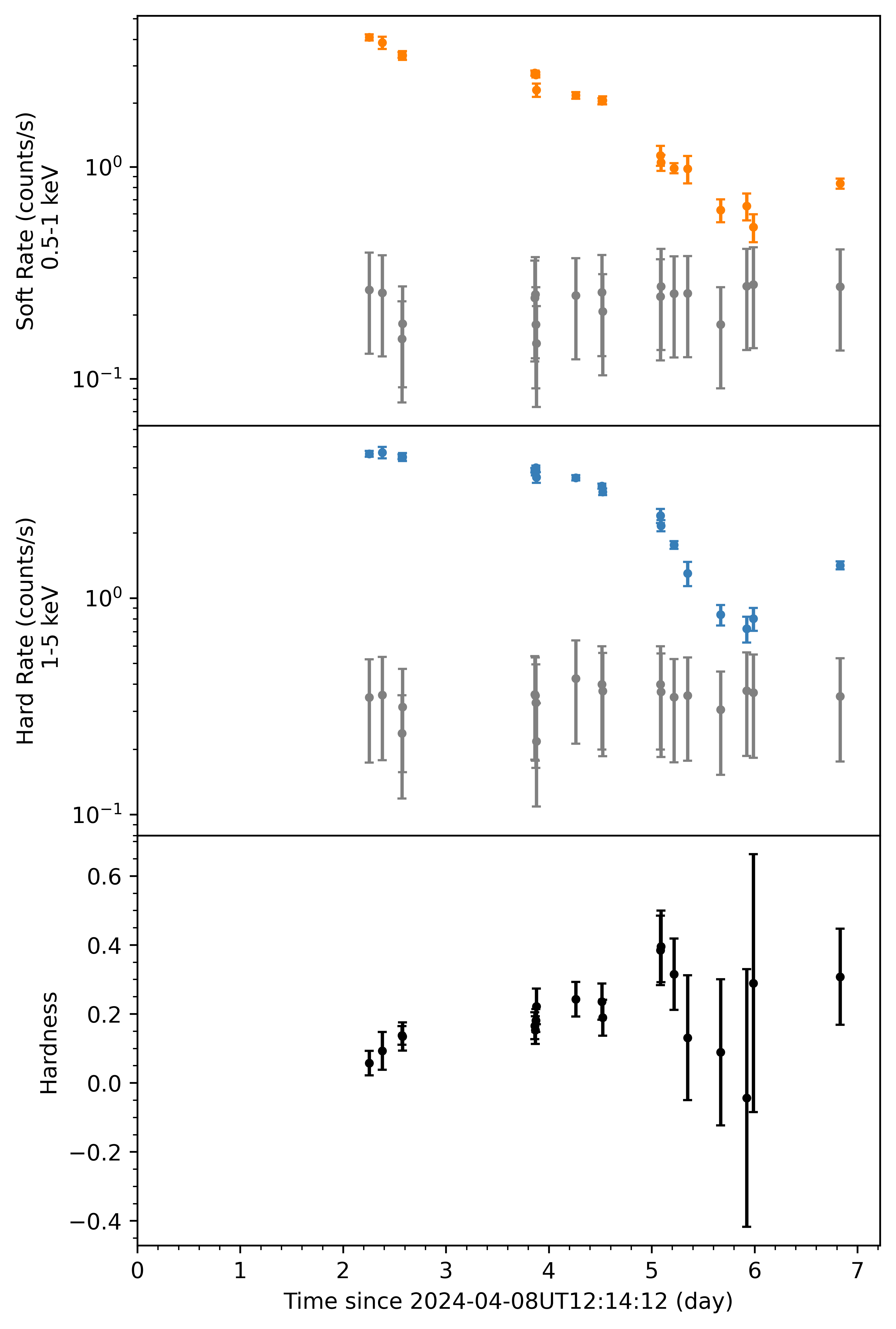}
\caption{NICER light curve of \src. The total count rates (i.e. without background subtraction) of the soft (0.5--1) and hard (1--5\nonbreak{}keV) bands are plotted in the upper and middle panels, respectively. In both panels, the background count rate in the respective band estimated with the SCORPEON model is plotted in gray dots. The hardness, defined as $(H-S)/(H+S)$ where $H$ and $S$ are the $net$ count rates of the soft and hard bands, respectively, is presented in the bottom panel.\label{fig:lc_nicer}}
\end{figure*}

We present the results from the \textit{EP}/WXT observations in Fig.~\ref{fig:lc_wxt} (right panel) and those from \textit{NICER}/XTI in Fig.~\ref{fig:lc_nicer}.
Overall, the evolution of the count rates shows little variation from the beginning up to $\sim$3--4\nonbreak{}days later, followed by a fast drop down to a level that is one order of magnitude lower around day 6.
Interestingly, the count rate shows somewhat increase again on day 7 as measured by \textit{NICER} (ObsID 7204340106).
\src{} was further monitored with \textit{Swift}/XRT and \textit{NICER}/XTI later in April, 2024; no X-ray emission has been detected ever since, however, indicating that its X-ray emission has likely faded away. 

The spectral shape exhibits a gradual and slight hardening over time until day 5, thereafter followed by spectral softening within the next day, up to day 6. However, the uncertainties are too large to draw a firm conclusion.

We also present intra-observation count rates and hardness ratios during the WXT observation of the initial source detection (left panel in Fig.~\ref{fig:lc_wxt}).
The observation is divided into three epochs: pre-flare, flare, and post-flare,
with the flare's time interval determined by the BB algorithm, as described in Sec.~\ref{sec:detection}. 
The count rate of the flare is more than two orders of magnitude larger than the persistent level, whereas the hardness ratio of the flare is statistically consistent with that of the persistent emission.
The result also confirms the aforementioned finding that X-ray emission from the source was already present even prior to the short-term flare.

\subsection{X-ray spectral property}

Here we present the analysis of the X-ray spectra of \src{}, based on the data acquired by \textit{EP}, \textit{NICER} and \textit{Swift}, as reduced in Sec.~\ref{sect:data}.
For the XRT and FXT spectra, the energy band of 0.5--10\nonbreak{}keV is selected, while the 0.5--4\nonbreak{}keV band is used for WXT.
For the XTI spectra, a broader energy band of 0.5--12\nonbreak{}keV is employed to better constrain the XTI background component. 
We perform X-ray spectral fitting with \texttt{Xspec} \cite{arnaud_xspec_1996} v12.14.1.
We group all the spectra with a minimum of 3 counts in each bin and the best fit is found by minimizing the Cash statistics \cite{cash_parameter_1979}, given the low number of counts of XTI above $\sim$10\nonbreak{}keV. 
The $wilms$ abundance table~\cite{wilms_absorption_2000} is used.
The spectra are fit with three simple models: an absorbed power-law model ($tbabs * powerlaw$ in \texttt{Xspec}), an absorbed thermal blackbody model ($tbabs * bbody$) and an absorbed disk blackbody model ($tbabs * diskbb$). All uncertainties derived from the spectral fitting are reported at the 90\% confidence level. 
 
We begin by performing a joint fit of the FXT spectra (ObsId 08500000058) and the \textit{NICER} spectrum (ObsId 7204340101), which were observed quasi-simultaneously. 
This results in joint energy spectra of the best quality among all the X-ray observations available for \src{}. 
Exclusively for the joint fit, we add a multiplicative constant as a scaling factor of the normalization parameter to account for calibration uncertainties between different instruments. The inter-calibration constant $\mathcal{C}$ is fixed to unity for FXT-A, while it is allowed to vary for FXT-B and XTI. 

\begin{figure*}
    \includegraphics[width=\textwidth]{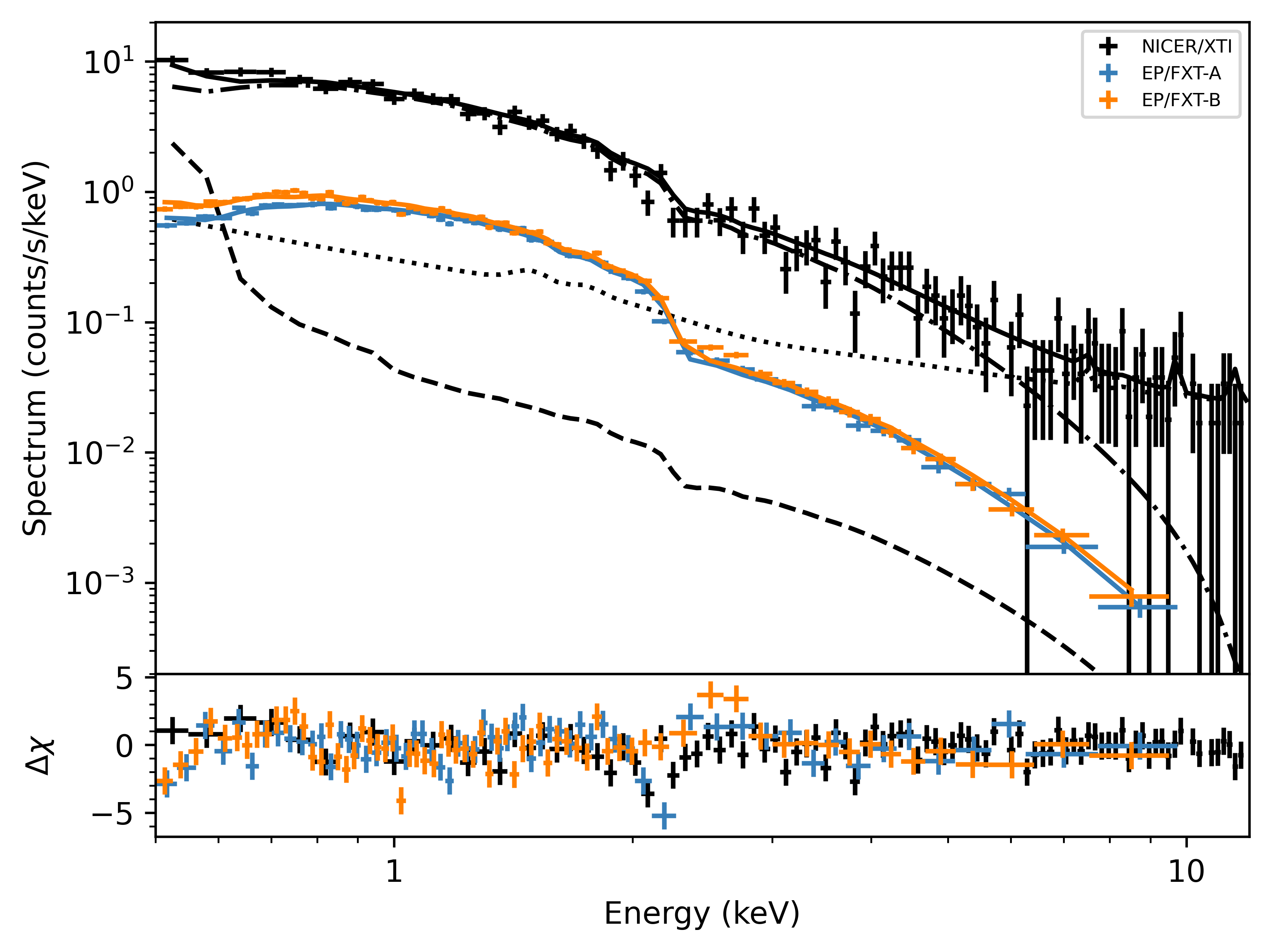}
    \caption{Upper: the energy spectra and the best-fit model for the joint XTI/FXT fit. The results of XTI, FXT-A, FXT-B are plotted in black, blue, and orange colors, respectively. The spectra of FXT-A and FXT-B are rebinned for clarity. For each instrument, the spectrum is represented by the crosses, while the best-fit model is plotted in a solid curve. For XTI, the source, sky background, and non-X-ray background models are plotted in dash-dotted, dashed, and dotted lines, respectively. Lower: $\Delta \chi$, defined as $(data - model) /error$. The residuals around $\sim$2.2\nonbreak{}keV for both XTI and FXT-A are probably due to imperfect calibrations of instrumental gold edges.\label{fig:fit}}
\end{figure*}

A power-law model provides an acceptable fit, yielding a cstat value of 1246.28 with 1147 degrees of freedom (dof). The energy spectra and the best-fit model are presented in Fig.~\ref{fig:fit}. In this model, the inter-calibration constant is $0.97\pm 0.01$ for FXT-B and $1.07^{+0.06}_{-0.04}$ for XTI, demonstrating consistency of the calibration across these instruments. The best-fit power-law photon index is $\Gamma$ = $2.10\pm 0.03$, and the hydrogen column density, that accounts for interstellar absorption along the line of sight ($tbabs$), is $N_{\rm H}$=$(1.6\pm 0.1)\times 10^{21}~\rm cm^{-2}$. For comparison, the $N_{\rm H}$ in this direction obtained from the neutral hydrogen HI4PI H\textsc{i} survey~\cite{hi4pi_collaboration_hi4pi_2016} is $6.2\times 10^{20}~\rm cm^{-2}$. The best-fit parameters are summarized in Table\nonbreak{}\ref{tab:fit}. However, an absorbed thermal model, either blackbody or disk blackbody, does not provide an acceptable fit (cstat/dof = 9679.10/1148 and 2550.11/1148, respectively). We thus conclude that the X-ray radiation of \src{} is of non-thermal origin.

Then, we proceed to fit the remaining spectra with the absorbed power-law model exclusively. In fitting the WXT and XTI spectra, we find the $N_{\rm H}$ value to be poorly constrained, so we fix it at the best-fit value of $1.6\times 10^{21}~\rm cm^{-2}$, derived from the joint fit. For the XRT spectrum, the $N_{\rm H}$ value is freely fitted, resulting in a best-fit value of $(1.5\pm 0.4)\times10^{21}~\rm cm^{-2}$, which is consistent with the value obtained from the joint fit of the FXT and XTI spectra. A simple absorbed power-law model provides acceptable fits across all the remaining spectra, with cstat/dof values ranging between 0.8 and 1.2. The fitted parameters are presented in Table\nonbreak{}\ref{tab:fit}. 

We also fit the WXT spectrum of the short-term X-ray flare, which was extracted from the time interval of the flare (12.3 seconds) derived above in the WXT observation (ObsID 13600005155). The spectral shape cannot be constrained due to limited source counts. Interesting enough, the flare reaches a flux of a few times $10^{-9}~\rm erg~cm^{-2}~s^{-1}$, 300 times the averaged value of the persistent emission before and after the flare.  
The measured fluence of the flare is $5^{+12}_{-2}\times 10^{-8}~\rm erg~cm^{-2}$. 

\subsection{Evolution of the X-ray emission}

\begin{figure*}
    \centering
    \includegraphics[width=\textwidth]{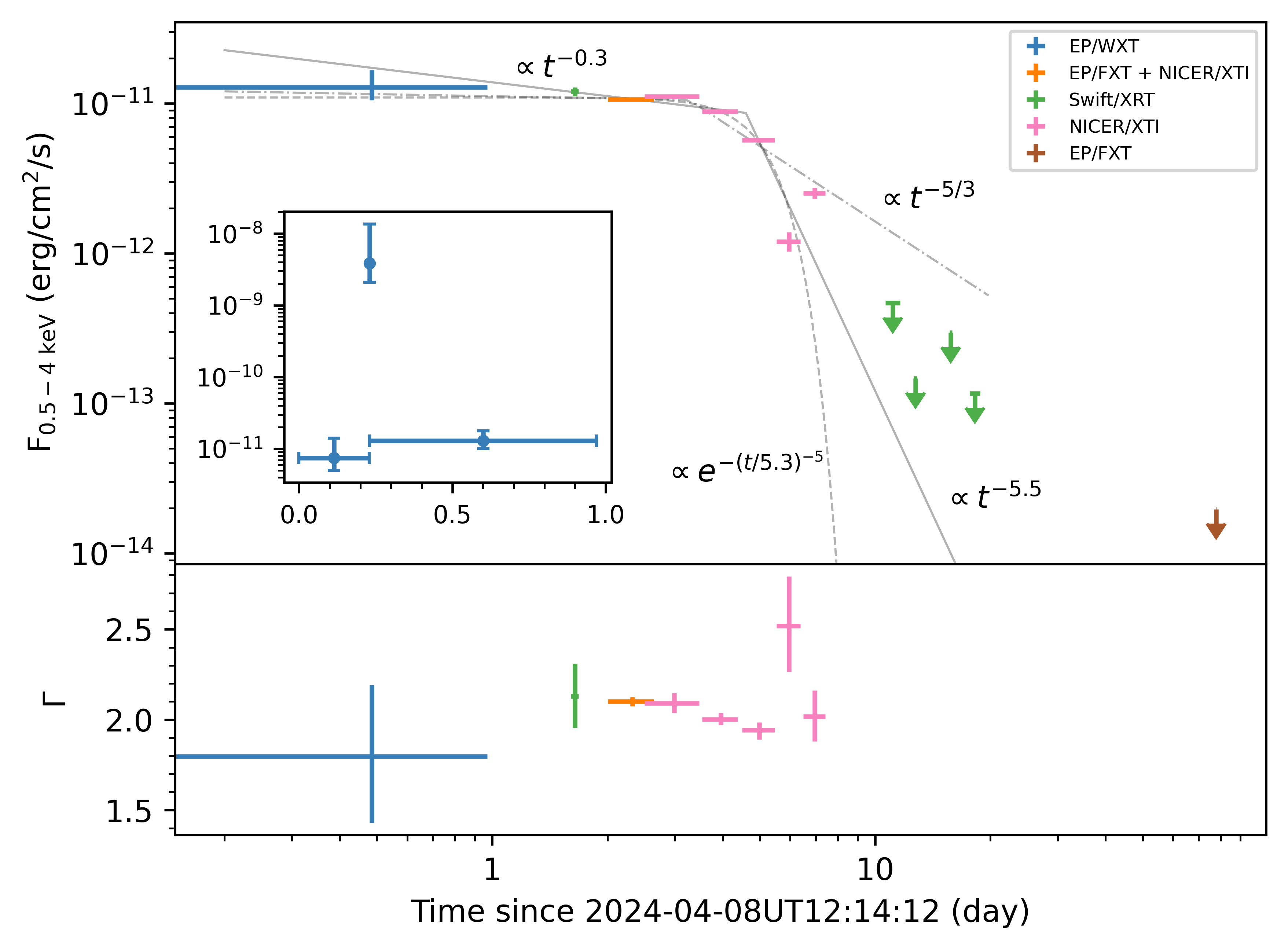}
    \caption{X-ray spectral evolution of \src. In the upper and lower panels, we present, as a function of time, the 0.5--4\nonbreak{}keV flux and the photon index, respectively. Results from various instruments are plotted in different colors, as indicated in the plot. The upper limits on the flux by \textit{Swift}/XRT observations taken after day 10 are plotted with green arrows, and the upper limit on the flux by the second FXT observation taken on June 25th is plotted with a brown arrow. In the inset plot of the upper panel, we also present the evolution of the flux within the first WXT observation. Gray curves represent phenomenological fits to the flux evolution: the best-fit broken power-law, and modified exponential decay (see text)
    are plotted in solid and dashed lines, respectively. The dash-dotted line represents the best-fit broken power-law model with the index after the break fixed at $-5/3$.
    \label{fig:spec}}
\end{figure*}

The temporal evolution of the best-fit values of the X-ray flux and power-law photon index is presented in Fig.~\ref{fig:spec} on logarithmic scales. 
To keep consistent with the WXT energy band,  we report the fluxes in the 0.5--4\nonbreak{}keV band for all instruments. 
Initially, the average X-ray flux, around $10^{-11}~\rm erg~cm^{-2}~s^{-1}$, appears to remain in a plateau phase with little or no decrease for $\sim$3--4\nonbreak{}days. This is followed by a steep decay to a flux level of $\sim$$10^{-12}~\rm erg~cm^{-2}~s^{-1}$ around day 6--7. 
Afterward, \src{} was no longer detected in X-rays during several \textit{Swift}/XRT observations taken $\sim$10\nonbreak{}days after the initial WXT detection. 
We obtain $3\sigma$ upper limits on the count rate at the source position utilizing \texttt{ximage} in the FTOOLS package, corresponding to flux limits of a few times $10^{-13}~\rm erg~cm^{-2}~s^{-1}$ in 0.5$-$4 keV. The flux limits are derived from the count rate to flux conversion using \texttt{WebPIMMS}\footnote{https://heasarc.gsfc.nasa.gov/Tools/w3pimms\_help.html}, assuming an absorbed power-law model with $N_{\rm H}=1.6\times 10^{21}~\rm cm^{-2}$ and $\Gamma=2$.
The upper limits set by \textit{Swift}/XRT are consistent with the steep decline in flux observed by \textit{NICER}/XTI. We also derive a $3\sigma$ upper limit on the flux in 0.5--4\nonbreak{}keV of $1.97\times 10^{-14}~\rm erg~cm^{-2}~s^{-1}$ for the second FXT observation taken on June 25th, based on the upper limits on the count rate (see Sec.~\ref{sec:data}) and the conversion factor from spectral fitting.

\begin{figure*}
    \includegraphics[width=\textwidth]{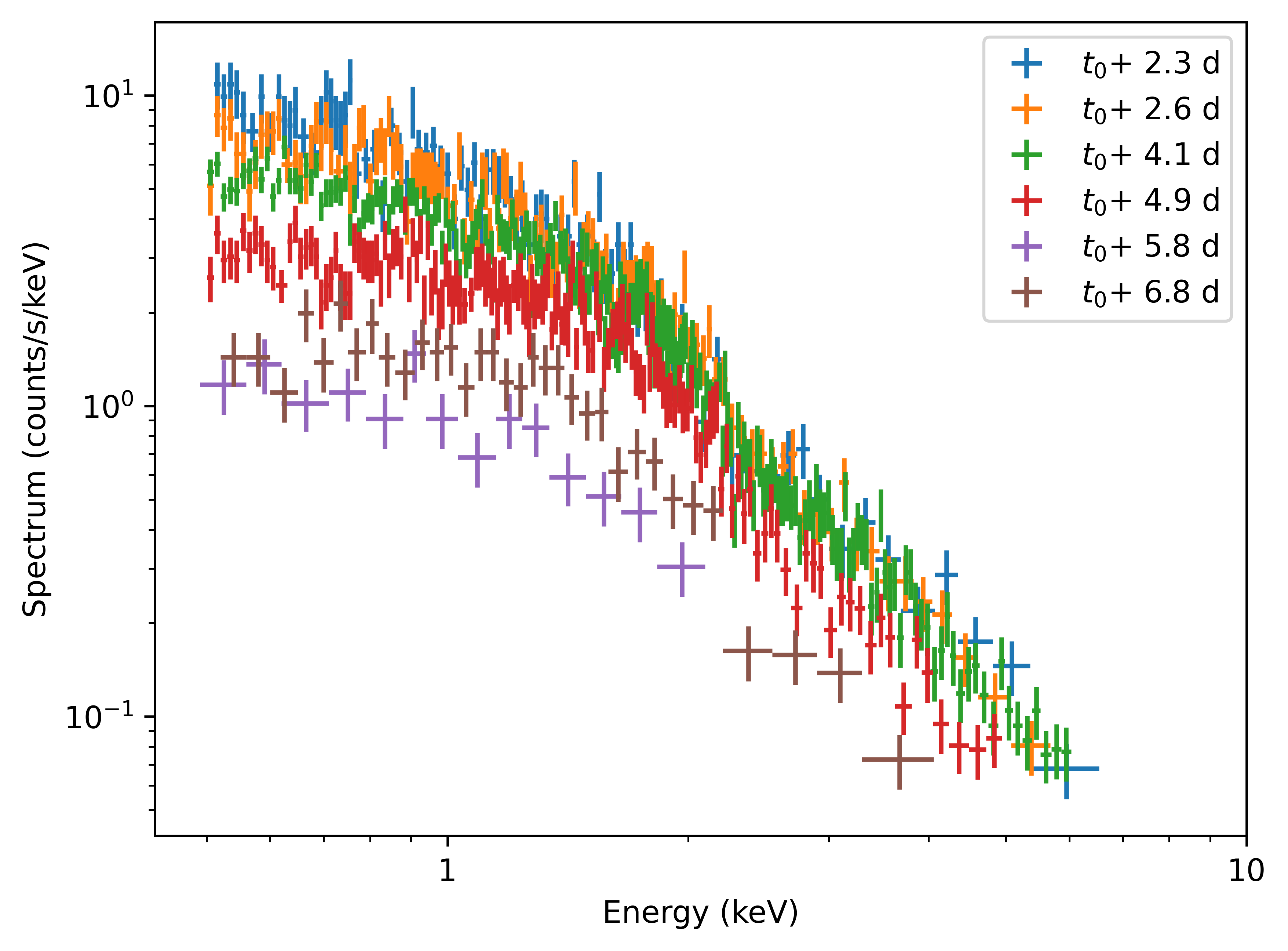}
      \caption{
    The evolution of the XTI spectra. For clarity, we rebin the spectra to ensure a minimum counts of 25 per bin. For each spectrum, we plot the data up to a critical energy above which the spectrum is dominated by the background. The critical energy varies from observation to observation and lies in the range of $\sim$2--6\nonbreak{}keV.
    \label{fig:specevo}}
\end{figure*}

\begin{table*}
\caption{Best-fit parameters of the absorbed power-law model fitted to X-ray spectra.\label{tab:fit}}
\centering
\begin{tabular}{lcccccccc}
\hline
\hline
$t-t_0$ & Inst.& ObsID & Exp. & $N_{\rm H}$ & $\Gamma$ & $F_{\rm 0.5-4~\rm keV}$ & Cstat/dof & Note\\
$\rm[day]$ & & & [ks] & $[10^{21}~\rm cm^{-2}]$	& &
$[10^{-11}~\rm cgs]$ &
& \\
\hline
0.0 & WXT & 155 & 38.5 & $1.6^*$ & $1.80^{+0.40}_{-0.37}$ & $1.28^{+0.39}_{-0.23}$ & 145.15/170 & \\
0.0 & WXT & 155 & 11.9 &  $1.6^*$ & $2.29^{+1.37}_{-1.09}$ &  $0.74^{+0.67}_{-0.24}$ & 78.34/67 & pre-flare \\
0.2 & WXT & 155 & $1.2\times 10^{-2}$ & $1.6^*$ &  $0.93^{+1.56}_{-1.49}$  & $385.38^{+977.62}_{-176.88}$ & 0.20/5 & flare \\
0.2 & WXT & 155 & 26.6 & $1.6^*$ & $1.86^{+0.45}_{-0.42}$ &  $1.29^{+0.49}_{-0.28}$ & 97.44/134 & post-flare \\
1.6 & XRT & 001 & 1.8 & $1.5\pm 0.4$ & $2.13^{+0.18}_{-0.17}$ & $1.20\pm{0.08}$ & 25.93/31 & \\
2.0 & FXT+XTI & 058/101 & 24.0/0.3 & $1.6\pm0.1$ & $2.10\pm0.03$ & $1.07\pm0.01$ & 1246.28/1147 & joint fit\\
2.6 & XTI & 102 & 0.5 & $1.6^*$ & $2.09^{+0.06}_{-0.05}$   & $1.11\pm0.04$ & 356.23/334 & \\
3.9 & XTI & 103 & 2.0 & $1.6^*$ & $2.00^{+0.04}_{-0.03}$   & $0.89\pm0.02$ & 703.17/565 & \\
4.5 & XTI & 104 & 1.4 & $1.6^*$ & $1.94^{+0.04}_{-0.05}$  &  $0.57\pm0.02$ & 558.78/460 & \\
5.7 & XTI & 105 & 0.3 & $1.6^*$ & $2.52^{+0.27}_{-0.25}$ &  $0.12\pm0.02$ & 134.97/127 & \\
6.8 & XTI & 106 & 0.5 & $1.6^*$ & $2.02\pm0.14$ & $0.25\pm0.02$ & 203.09/203 & \\
\hline
\hline
\end{tabular}
Notes. columns from left to right: the start time of the observation with respect to $t_0$=2024-04-08UT12:14:12, in day; the instrument; the observation ID; the exposure time in ks; the neutral hydrogen column density; the photon index; the flux of the 0.5--4\nonbreak{}keV band, in $10^{-11}~\rm erg~cm^{-2}~s^{-1}$; cstat/dof. A superscript asterisk indicates that the parameter is fixed in the spectral fitting. All uncertainties quoted correspond to the $90\%$ confidence level.
\end{table*}

We parameterize the 0.5--4\nonbreak{}keV flux (excluding upper limits) as a function of time since $t_0$ with a few phenomenological models. Fitting a broken power-law model to the data indicates that the shallow and steep decay phases are best described by $t^{-0.3}$ and $t^{-5.5}$, respectively, with the transition occurring at $t\sim 4.6~\rm days$ after trigger. The best-fit model is represented by a solid gray line in the upper panel of Fig.~\ref{fig:spec}. As can be seen in the figure, while the XRT upper limits are not included in the fit, the best-fit broken power-law model is well consistent with the upper limits. 
The decay index of the shallow decay phase is close to $0$, consistent with being a plateau phase. We experiment with fixing the index in the steep decay phase of the broken power-law at $-5/3$, the canonical TDE decay rate, and find the late-time emission to be obviously over-estimated (dash-dotted gray line). 
Alternatively, we use a modified exponential decay model that has the form of $F_{\rm X}\propto e^{-(t/\tau)^\alpha}$, where $F_{\rm X}$ is the X-ray flux and $t$ is the time since the first detection in days. This model (dashed gray line) seems to be roughly consistent with the data as well, yielding $F\propto e^{-(t/5.3)^{-5.0}}$, with the critical time $\tau=5.3~\rm days$, which is close to the transition time determined by the broken power-law fit. Furthermore, the Band function~\cite{band_batse_1993}, a smoothly broken power-law model, is also employed to fit the flux evolution. However, it does not fit the data adequately, 
indicating that the transition from the plateau to the steep decay phase is indeed sharp.

The best-fit values of the photon index varied around 2 with time. The evolution of the photon index exhibits a trend that is consistent with the XTI hardness ratio (bottom panel of Fig.~\ref{fig:lc_nicer}), as expected. 
The spectrum hardens during the shallow decay phase, and becomes softer in the observation 7204340101 that was taken $\sim 6$ days after the trigger. We also present the evolution of the XTI spectrum in Fig.~\ref{fig:specevo}, where an apparent hardening of the spectrum with time can be seen in the first 4 observations. The best-fit parameters of the absorbed power-law model are summarised in Table~\ref{tab:fit}.

\subsection{Summary of the observed properties}
Here we summarize the observed properties of \src{}:
\begin{itemize}
    \item \src\ was initially discovered as an X-ray transient, which was persistent at a flux level of $\sim$$10^{-11}~\rm erg~cm^{-2}~s^{-1}$ in 0.5--4\nonbreak{}keV throughout a 83.8\nonbreak{}ks-long observation. 
    An intense soft X-ray flare lasting only for $\sim$12\nonbreak{}s was also detected, reaching $3.9\times 10^{-9}~\rm erg~cm^{-2}~s^{-1}$ or $\sim$300 times the persistent X-ray flux. The flare exhibits three peaks, each of a roughly symmetric profile.
    No other short-timescale variations than the fast flare were found during this observation.  
    
    \item Further monitoring observations with \textit{EP}, \textit{NICER} and \textit{Swift} reveal that the X-ray emission of \src\ stayed in a plateau phase that lasts for $\sim $4.6\nonbreak{}days, followed by a steep decay with time as $\propto t^{-5.5}$. The last significant detection of \src{} in X-rays was made by \textit{NICER}, approximately 7\nonbreak{}days after the trigger.
    After $\sim$10\nonbreak{}days, \src{} became no longer detectable in X-rays. 
    Upper limits can be imposed on its 0.5--4\nonbreak{}keV flux as a few times $ 10^{-13}~\rm erg~cm^{-2}~s^{-1}$ around 10--20\nonbreak{}days and $2\times 10^{-14}~\rm erg~cm^{-2}~s^{-1}$ two months after the initial detection. \src{} was undetected during the WXT observation taken $\sim$13\nonbreak{}days before the discovery. Hence, the duration of the X-ray emission is found to lie in the range of $\sim$7--23\nonbreak{}days, making \src\ an X-ray transient with a  intermediate timescale.
    \item The X-ray spectrum of \src{} is non-thermal and can be fit with an absorbed power-law model. The photon index varies with time, ranging from 1.8 to 2.5.
    \item No counterparts have been found in the optical, NIR and NUV bands, with the earliest observation started 17 hours after the initial X-ray detection and the latest 24 days after, setting upper limits (3\,$\sigma$) to the magnitudes of $\sim$21--22\,mag in the various bands. 
    \item No radio emission was detected at $\sim$30 days after the source discovery, with upper limits (5\,$\sigma$ upper) of $\sim$125 and $\sim$200\,$\mu$Jy to the fluxes in the C- and K-bands, respectively. 
    
\end{itemize}





\section{Discussion}
\label{sec:discussion}

Based on its temporal and spectral properties, \src\ seems to resemble none of the archetypes of the various classes of X-ray transients known so far.
The lack of multi-wavelength (in particular optical) counterparts and thus redshift measurement makes it even more difficult to identify its nature. 
Nevertheless, we try to gain insight into the nature of this peculiar source by comparing its properties with those of the known transient types that are possibly related to \src\ to some extent, namely TDE with relativistic jets, gamma-ray burst, X-ray binary and fast blue optical transient.   
We note that, in recent years, there have been findings of fast X-ray transients with timescales of a few kilo-seconds or even shorter, as reported for examples in \cite{irwin_ultraluminous_2016,bauer_new_2017,xue_magnetar-powered_2019}, whose origins are largely unclear although model interpretations have been proposed for some of them.
Obviously, the X-ray radiation from \src{} lasted too long to be possibly associated with any of them. 

\src{} was not detected in the optical or NIR band, even in the very early observations taken within two days after the X-ray trigger. To examine if this is intrinsic or due to strong extinction, we estimate the optical extinction $A_{\rm V}$ by using the relation between the neutral hydrogen column density $N_{\rm H}$ and $A_{\rm V}$~\cite{guver_relation_2009}, and find $A_{\rm V} \simeq 0.7$\nonbreak{}mag by taking $N_{\rm H}=1.6\times 10^{21}~\rm cm^{-2}$ as obtained from the X-ray spectral fitting. The value of $A_{\rm V}$ suggests of merely a factor of two decrease in the V-band flux due to extinction. The extinctions in the longer wavelength J, H, K$_s$, and r-bands are expected to be even lower, as the extinction monotonically decreases with the wavelength in the optical/NIR band. Therefore, we conclude that \src{} is intrinsically weak in optical and NIR.

Before proceeding with detailed comparison and discussion, we can constrain the size of the X-ray emitting region from the timescale of the fastest variability, regardless of the physical process of the X-ray emission, as
 $R_{\rm em} \lesssim \delta c t_{\rm var} / (1+z)$, where $R_{\rm cm}$ is the characteristic co-moving size of the emitting region, $z$ the redshift, $\delta$ the Doppler factor, $c$ the speed of light and $t_{\rm var}$ the observed timescale of the fastest variability. 
 Adopting the timescale of the intense short-term flare of $t_{\rm var}\sim 12$\nonbreak{}s, we can set a limit on the size of the X-ray emitting region (as measured in the co-moving frame of the emitter) in \src\ as:
\begin{equation}
\label{eq:rem1}
R_{\rm em} \lesssim 3.6\times 10^{11}\frac{\delta}{1+z} ~\rm cm.
\end{equation}
If \src{} is powered by a black hole with mass $M$: 
\begin{equation}
\label{eq:rem}
R_{\rm em} \lesssim 1.2\times 10^6\frac{\delta}{1+z} \frac{M_\odot}{M}R_{\rm S},
\end{equation}
where $R_{\rm S}\equiv 2GM/c^2$ is the Schwarzschild radius of the black hole with $G$ being the gravitational constant.

\subsection{Jetted tidal disruption event}
\src{} has distinct X-ray spectral properties compared to thermal TDEs that are detected in X-ray. Thermal TDEs usually show supersoft thermal X-ray spectra with blackbody temperatures of a few tens of eV (e.g., refs.\cite{saxton_x-ray_2020,gezari_tidal_2021}), especially at the early stage of the outburst. In contrast, \src{} has a clearly non-thermal spectrum, with a photon index of $\sim$2.

On the other hand, the known jetted TDEs have non-thermal, hard X-ray spectrum,  similar to that of \src. The radiation of jetted TDEs is believed to originate from a collimated relativistic jet formed after the tidal disruption process. We compare \src{} with Sw J1644+57, the prototype jetted TDE and the one with the most extensive dataset among all jetted TDEs known so far.
We find the two sharing the following similarities: i) the evolution of the X-rays from Sw J1644+57 also exhibits a plateau phase that lasted for $\sim$10\nonbreak{}days, followed by a power-law decay~\cite{tchekhovskoy_swift_2014}. An additional feature in the X-ray light curve of Sw J1644+57 is the extreme flaring activity that is present in the early plateau phase, possibly due to precession or wobbling of the relativistic jet~\cite{tchekhovskoy_swift_2014}. The short-term flare found in \src\ may also be explained in the same way, and its much shorter duration than that of Sw J1644+57 may be ascribed to that the bulk of the flare emission is below the moderate sensitivity of EP/WXT. 
ii) For Sw J1644+57, no transient optical emission was found in the early observations, suggestive of intrinsically weak optical emission produced (e.g., refs.\cite{bloom_possible_2011,levan_extremely_2011}). 
The optical property of \src\ is consistent with this picture. 

\begin{figure*}
    \includegraphics{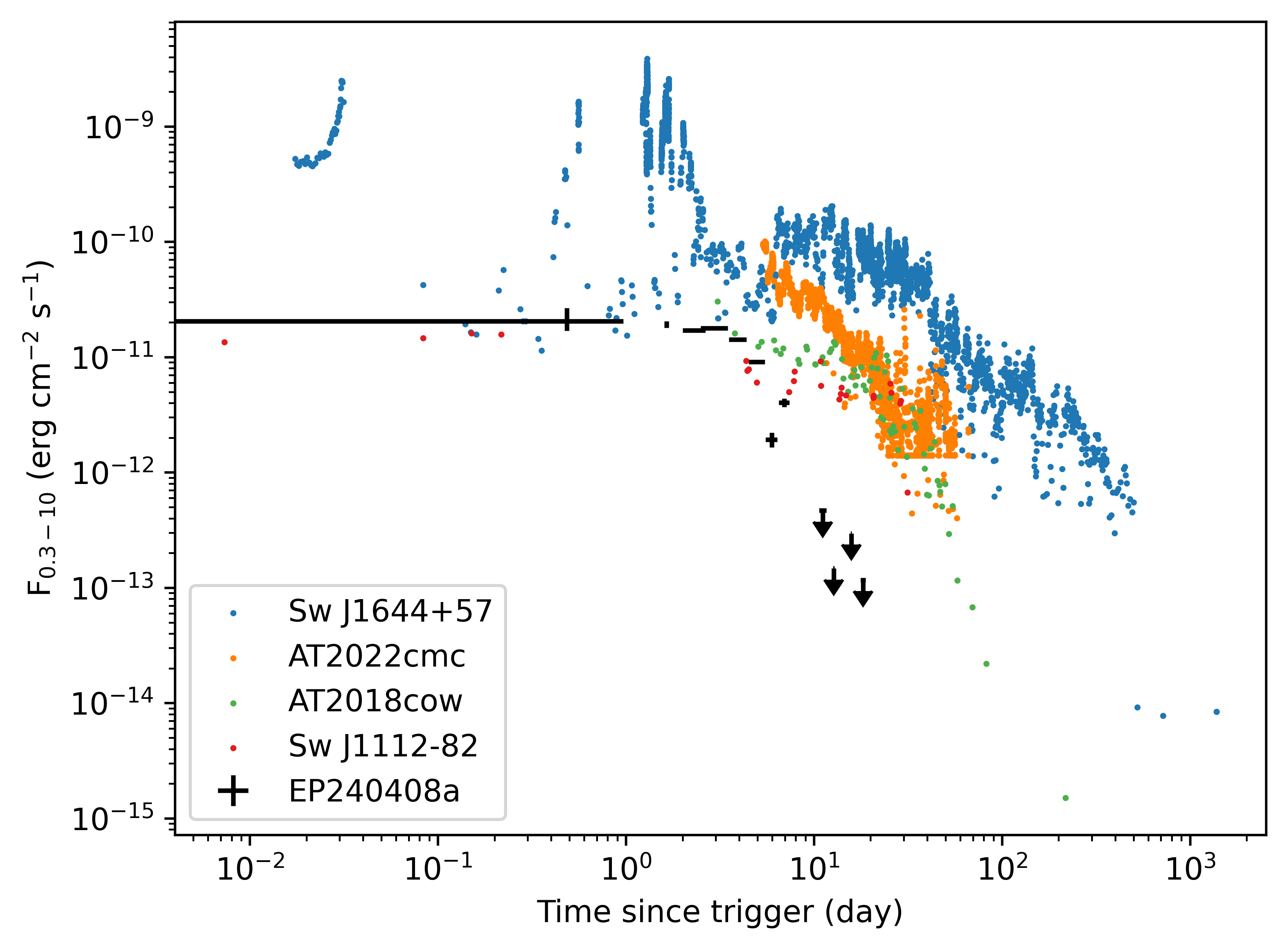}
    \caption{Comparison of \src{} with Sw J1644+57 (the prototype jetted TDE; blue dots), AT2022cmc (a jetted TDE; orange dots), Sw J1112-82 (a jetted TDE; red dots), and AT2018cow (an FBOT; orange dots) on their long-term X-ray light curves. The light curves of Sw J1644+57 and AT2022cmc are taken from refs.~\cite{polzin_luminosity_2023}. The light curve of AT2022cmc consists measurements by \textit{NICER} (taken from refs.~\cite{pasham_birth_2023}) and \textit{Swift}, and for the latter we obtain the light curve using the \textit{Swift} online product generator and convert the count rate to flux utilizing WebPIMMS by assuming an X-ray spectra model that is obtained from refs.~\cite{yao_-axis_2024}. The XRT count rate of Sw J1112-82 are also taken from the \textit{Swift} online product generator and converted to 0.3--10\nonbreak{}keV flux using WebPIMMS assuming the X-ray spectrum obtained from refs.~\cite{brown_swift_2015}. For \src{}, we extrapolate its X-ray flux from the 0.5--4\nonbreak{}keV band to the 0.3--10\nonbreak{}keV band, assuming the absorbed power-law model with $\Gamma=2$ and $N_{\rm H}=1.6\times 10^{21}~\rm cm^{-2}$.\label{fig:tde}}
\end{figure*}

\begin{figure*}
    \includegraphics{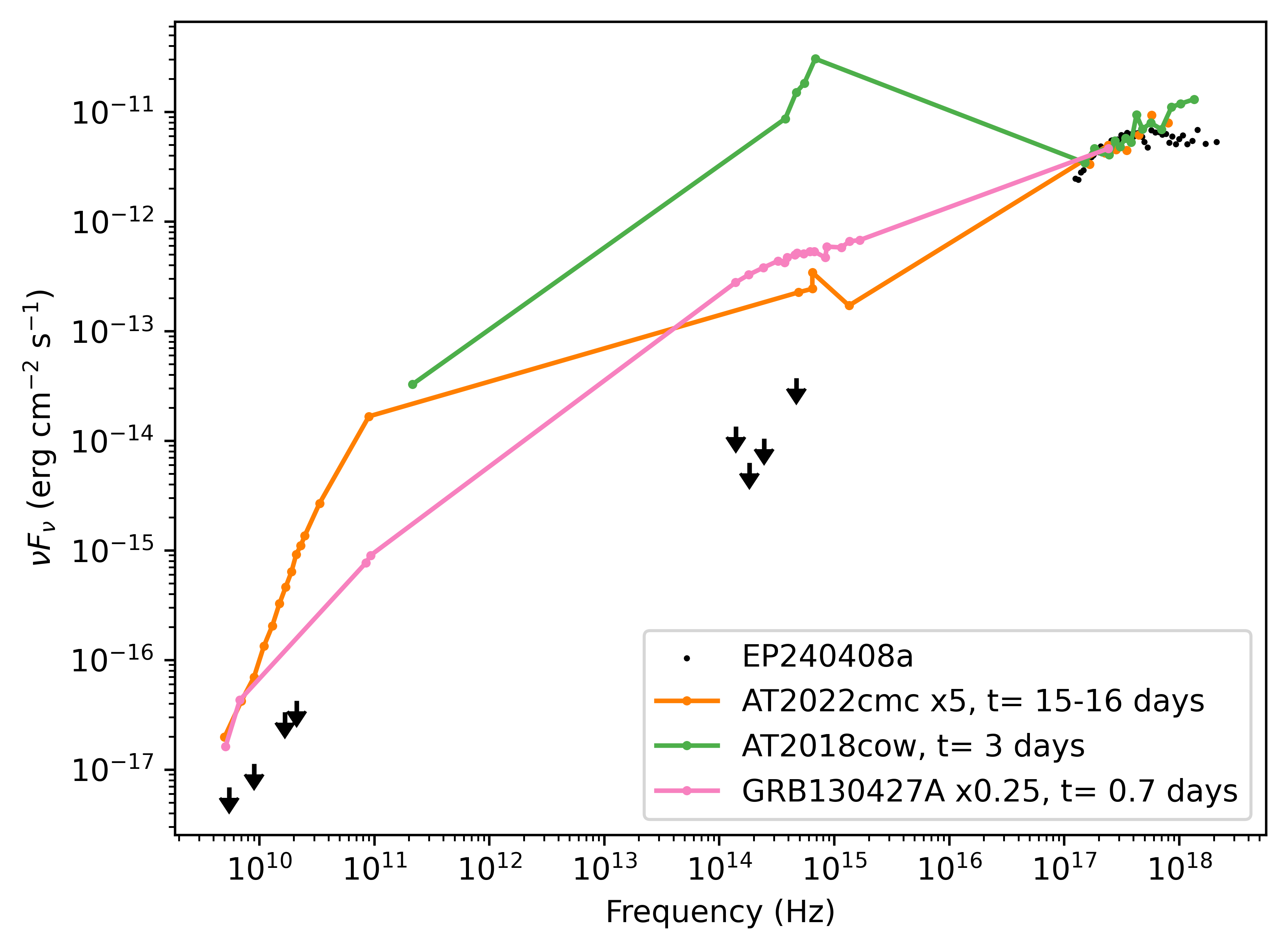}
    \caption{Comparison of \src{} (black) with AT2022cmc (a jetted TDE; orange), AT2018cow (an FBOT; orange), and GRB130427A (a GRB; pink) on their broadband SED. Data sources: AT2022cmc~\cite{pasham_birth_2023}; AT2018cow~\cite{margutti_embedded_2019}; GRB130427A~\cite{perley_afterglow_2014}. For AT2022cmc and GRB130427A, the SED is shifted to match \src{} in the soft X-ray band, as indicated in the plot. The data of \src{} consist of measurements made with ATCA, GROND, GSP, and FXT-A, and the observations in different energy bands are not simultaneous. For each object other than \src{}, there are multiple epochs during which quasi-simultaneous radio, optical/IR, and soft X-ray measurements are available, and for constructing the SED we choose the earliest epoch.\label{fig:sed}}
\end{figure*}

In Fig.~\ref{fig:tde}, we compare the X-ray light curves between \src{} and three jetted TDEs currently known, namely, Sw\,J1644+57 (blue),  AT2022cmc~\cite{pasham_birth_2023} (orange) and Sw\,J1112-82~\cite{brown_swift_2015} (red). 
In the plateau phase, Sw\,J1644+57 is around one order of magnitude brighter than \src. A similar luminosity of the two at the plateau phase would therefore require the host galaxy of \src{} to be located at a higher redshift of $\sim$0.9 than that of Sw\,J1644+57 (0.35). We also compare the broadband spectral energy distribution (SED) between \src{} and AT2022cmc in Fig.~\ref{fig:sed}.

If the X-ray flare originated from a relativistic jet aligned close to the line of sight, the Doppler factor $\delta\sim \gamma$, where $\gamma$ is the Lorentz factor of the jet. The Lorentz factor of jetted TDEs is constrained to be in the range of $\sim$10--100~\cite{bloom_possible_2011,yao_-axis_2024}. Using equation~\ref{eq:rem1} and assuming $z=0.9$, we can place an upper limit on the size of the emitting region, as measured in the $host$ frame, as  $R^\prime_{\rm em} \lesssim 1.9\times 10^{13}-1.9\times 10^{17}~\rm cm$ (here we make use of the fact that the sizes of the emitting region measured in the co-moving and host frames are different by a factor of $\gamma$.) The constraint on the emitting size here is consistent with that of Sw\,J1644+57 [$R^\prime_{\rm em}\sim 2.3\times 10^{14}(\gamma/10)^2~\rm cm$; \cite{bloom_possible_2011}] and AT2022cmc [$R^\prime_{\rm em} \lesssim 1.2\times 10^{16} (\gamma/30)^2~\rm cm$; \cite{yao_-axis_2024}].

On the other hand, \src\ also shows some other properties that are difficult to explain in the jetted TDE scenario. 
While both \src{} and Sw J1644+57 follow a power-law decay after the plateau phase, the decay of \src{} is much steeper than Sw J1644+57, with the latter showing a canonical $\propto t^{-5/3}$ decay that is consistent with TDE theory. Actually, Sw J1644+57 does show a rapid decay $\sim$500\nonbreak{}days after trigger that was interpreted as the shutting off of the relativistic jet. 
If the steep decays of the two share the same physical origin, then a question arises as to why the jet in \src{} turns off much faster than the jet in Sw J1644+57. 
In the case of AT2022cmc, no plateau is present in the X-ray light curve, which follows simply a power-law decay $\propto t^{-2}$~\cite{yao_-axis_2024}. 

The TDE scenario for \src\ is further challenged by its weak low-frequency radio and IR emissions observed, in contrast to Sw J1644+57 and AT2022cmc, both showing bright radio emission and IR brightening.
The low-frequency synchrotron radiation is expected to be produced from relativistic jets. It is known that the radio emission of jetted TDEs peaks at a much later time than X-rays, i.e. 150 -- 200\nonbreak{}days later in the case of Sw J1644+57~\cite{zauderer_radio_2013}. 
We estimate the radio flux of \src\ at $\sim$30\nonbreak{}days after the trigger by scaling it with the flux of Sw J1644+57 ($\sim$20\nonbreak{}mJy at $\sim$20\nonbreak{}GHz) \cite{zauderer_birth_2011} using the flux ratio of the X-ray plateaus (one order of magnitude higher in Sw J1644+57 than in EP240408a).
The estimated flux at $\sim$20\nonbreak{}GHz is $\sim$2\nonbreak{}mJy,  
which is much higher than the upper limit of $\sim$200\nonbreak{}$\mu$Jy derived from the ATCA observations. 
The deficit of strong radiation at the low frequencies in \src{} can be clearly seen in Fig.~\ref{fig:sed} compared to AT2022cmc, which shows much brighter emission in the radio and optical/NIR bands. 
The weak optical/NIR emission in \src{} cannot be explained by extinction, given the relatively small value of $A_{\rm V} \simeq 0.7$\nonbreak{}mag along its line of sight.


These two issues with the jetted TDE interpretation may be mitigated if the TDE was produced by an intermediate-mass black hole (IMBH). First, the debris fallback timescale $t_{\rm fb} \simeq 4~M_4^{1/2} r_*^{3/2} m_*^{-1}$ days, is much shorter for a lower mass BH than for a higher mass one. This could be in line with the observed early onset of the rapid X-ray decay, if the jet luminosity evolution is tied to that of the debris mass fallback rate. Secondly, one of the most preferable sites that harbors an IMBH is the core of a globular star cluster (e.g., refs~\cite{lin_luminous_2018}), which are known to lack circum-centre medium. This may explain the non-detection of radio emission at an early time $t \sim 30$ day in EP240408a, since the jet has not yet shock-swept sufficient medium to produce bright synchrotron emission. However, this jetted IMBH-TDE interpretation remains speculative due to the scarcity of the multi-wavelength data.  

\subsection{Gamma-ray burst}
The bright, short-term flare with a duration of $12.3~\rm s$ is reminiscent of the prompt emission of long gamma-ray bursts.
However, the GRB interpretation is challenged by the detection of the relatively stable persistent X-ray emission before and after the flare. Although emission before the main prompt emission, known as precursors, have been observed in both long and short GRBs, the precursor emission is usually a few seconds long with a separation of a few to tens of seconds \cite{burlon_precursors_2008,troja_precursors_2010,hu_internal_2014}, which is clearly different from the persistent, steady emission detected in \src{}. GRB X-ray afterglow light curves typically have a plateau extending to $10^3-10^4$ seconds followed by a decay with a slope of $\sim t^{-1}$ \cite{zhang_physical_2006}. This is also very different from the observed behavior: the duration of the plateau is much longer and the slope of the post-plateau phase much steeper. \src{} and GRBs are distinct in their SED as well; for instance, strong radio and optical emission is observed in the afterglow of GRB130427A, as shown in Fig.~\ref{fig:sed}. We therefore conclude that the source is very unlikely to be a GRB.

We note that, EP240315a, a GRB that was also detected by \textit{EP}, exhibits a few interesting features in the soft X-ray band. Its prompt emission in the soft X-ray band (0.5--4\nonbreak{}keV) has a significantly longer duration than the prompt emission in the $\gamma$-ray band, with a $T_{\rm 90}$ of $\sim$1,000\nonbreak{}s in the 0.5--4\nonbreak{}keV band compared with a $T_{90}$ of $\sim$40\nonbreak{}s in the $\gamma$-ray band~\cite{liu_soft_2024}. The trigger in the soft X-ray band is earlier than that in the $\gamma$-ray band. Compared with EP240315a, the duration of the steady soft X-ray emission before the short flare is much longer ($\gtrsim 20~\rm ks$). The behaviors of EP240315a in terms of the transition time from the plateau to the power-law decay and the index of the power-law are similar with most GRBs~\cite{liu_soft_2024} and are therefore different with \src{}.

\subsection{X-ray binary}
Another possible scenario is that \src{} may be an X-ray binary (XRB). XRBs are bright X-ray emitters. In this case, the short-term flare may be a type-I thermonuclear burst that is often seen in neutron star X-ray binaries (NSXRBs). The duration of the normal type-I bursts (in contrast to the intermediate duration bursts or superbursts, both of which are more energetic and have longer durations) is in the range of 10--100\nonbreak{}s (e.g., refs.~\cite{galloway_thermonuclear_2021}), consistent with the short-term flare seen in \src. Assuming that the flare had reached the Eddington luminosity of a neutron star, the distance to the source would be $\sim$18\nonbreak{}kpc, leading to the X-ray luminosity of \src{} during the plateau phase to be a few times $10^{35}~\rm erg~s^{-1}$. However, the profile of the flare is distinct from those of the type-I bursts that typically exhibit a fast rise and an exponential decay. 
Though the possibility of being an XRB cannot be ruled out, several observed properties of \src{} seem to be different from those of typical XRBs. \src{} decays rapidly after the plateau phase within a few days, while the outbursts of XRBs take longer time to decay. Also, XRBs are known to be multi-wavelength emitters, and the intrinsic weakness of \src{} in the radio, IR, and optical bands would be puzzling if it is an XRB.

\subsection{Fast blue optical transient}
We also compare \src{} with the fast blue optical transient (FBOT) AT2018cow \cite{prentice_cow_2018}. The shapes of the X-ray light curves are quite similar, both with a plateau transitioning to a steep decay. 
The transition timescale (from plateau to steep decay) of AT2018cow is a bit longer than \src. While the X-ray fluxes of the two during the plateau phase are similar (both at $\sim$$ 10^{-11}~\rm erg~cm^{-2}~s^{-1}$), they are drastically different in the optical brightness. AT2018cow has a peak brightness of $13.8$\,mag in the $g^\prime$-band \cite{prentice_cow_2018}, at least 6 magnitudes brighter than \src{}, as evidenced in Fig.~\ref{fig:sed} where a comparison of their SEDs is shown. The sharp contrast makes the FBOT scenario for \src{} unlikely.

\section{Summary}
\label{sec:conclusion}
We report the discovery of a new X-ray transient, EP240408a, by \textit{EP}/WXT on April 8th, 2024 and follow-up characterization of the source properties by observations made with space- (\textit{NICER}, \textit{Swift} and \textit{EP}) and ground-based telescopes (including GROND, ATCA, NOT) at multi-wavelengths. 
Remarkably, a short (lasting for $\sim$12\nonbreak{}s) yet intense flare occurred once in its X-ray light curve during the $83.8$\nonbreak{}ks-long WXT observation in which the transient was first detected.
The flare reached a flux of $3.9\times 10^{-9}~\rm erg~cm^{-2}~s^{-1}$, $\sim$300 times brighter than the persistent flux level of $\sim$$10^{-11}~\rm erg~cm^{-2}~cm^{-1}$ in the 0.5--4\nonbreak{}keV band.

All the X-ray spectra can be well fitted with an absorbed power-law model, with the photon index varying with time and falling in a range of $\sim$1.8--2.5. 
The X-ray flux appeared to be relatively stable in a plateau phase lasting for ${\rm }~5$ days after the trigger, and thereafter decayed rapidly to levels of being undetectable $\sim$10 days after the initial detection ($5\sigma$ upper limits of a few times $10^{-13}~\rm erg~cm^{-2}~s^{-1}$). 
The characteristic duration of the transient X-ray emission is estimated to lie in between 7--23\nonbreak{}days. 
This makes \src\ an X-ray transient with an intermediate timescale, in contrast to fast transients (with timescales less than a day or so) and long-term transients (from months to years) that are more commonly found.

No counterparts in the optical, NUV and NIR bands have been found, with limits of around $\sim $20\nonbreak{}mag set in various optical/NIR bands starting from within one day after the initial X-ray detection. Given the low optical extinction of $A_{\rm V}\simeq 0.7$\nonbreak{}mag, the weakness of the emission in the optical/NIR bands is suggested to be intrinsic. No radio counterparts have been found either based on the ATCA C- and K-band observations performed $\sim$30 days after the first X-ray detection.

The observed properties and possible origin of \src{} are discussed. 
While \src{} shares some of the behaviors similar, to some extent, to several well-known types of X-ray transients known so far, including jetted-TDEs, GRBs, FBOTs, and XRBs, we demonstrate that its unique temporal and spectral properties cannot be readily explained by any of these types.
Thus, its nature remains an enigma. 
We suggest that \src{} may represent a new type of transients with intermediate timescales of the order of $\sim$10\nonbreak{}days, that may have been missed in previous time-domain X-ray surveys. 
Any constraints on the redshift of \src{} would be valuable, which is challenging to measure due to the faintness of its presumed optical counterpart.

EP240408a was discovered two months after \textit{EP}-WXT in operation. Since then no other similar sources have been found in the following six months so far. The \textit{EP} detection rate for similar sources is estimated to be likely 1--2 events per year. Given the X-ray flux levels and timescale of EP240408a, it would be difficult to detect similar objects by other X-ray telescopes or telescopes in other wavebands currently operating. Therefore, it is not surprising that this kind of transients was first discovered by \textit{EP}. 
Therefore, this discovery is likely to reveal the existence of a potentially new population of transients. \textit{EP} and other future X-ray monitoring missions are expected to detect more of such transients promptly. 
Meanwhile, timely and deep follow-up observations at multi-wavelengths and measurements of the redshifts of future events are essential for further characterizing their properties and to understand their physical origins.


\Acknowledgements{The authors acknowledge the referee for his/her prompt and valuable comments. This work is based on data obtained with Einstein Probe, a space mission supported by Strategic Priority Program on Space Science of Chinese Academy of Sciences, in collaboration with ESA, MPE and CNES (Grant No. XDA15310000), the Strategic Priority Research Program of the Chinese Academy of Sciences (Grant No. XDB0550200), and the National Key R\&D Program of China (2022YFF0711500). This work is also based on data obtained by the Neil Gehrels Swift Observatory (a NASA/UK/ASI mission) supplied by the UK Swift Science Data Centre at the University of Leicester, as well as on data obtained with NICER, a $0.2-12$ keV X-ray telescope operating on the International Space Station, funded by NASA. The Australia Telescope Compact Array is part of the Australia Telescope National Facility (https://ror.org/05qajvd42) which is funded by the Australian Government for operation as a National Facility managed by CSIRO.
We acknowledge the support by the National Natural Science Foundation of China (Grant Nos. 12333004, 12321003, 12103065, 12373040, 12021003, 12025303, 12393814, 12203071), the China Manned Space Project (Grant Nos. CMS-CSST-2021-A13, CMS-CSST-2021-B11), and the Youth Innovation Promotion Association of the Chinese Academy of Sciences. F.C.Z is supported by a Ram\'on y Cajal fellowship (grant agreement RYC2021-030888-I). L.G. acknowledges financial support from AGAUR, CSIC, MCIN and AEI 10.13039/501100011033 under projects PID2023-151307NB-I00, PIE 20215AT016, CEX2020-001058-M, and 2021-SGR-01270. We acknowledge the data resources and technical support provided by the China National Astronomical Data Center, the Astronomical Science Data Center of the Chinese Academy of Sciences, and the Chinese Virtual Observatory.
Part of the funding for GROND (both hardware and personnel) was generously granted from the Leibniz-Prize to G. Hasinger (DFG grant HA 1850/28-1) and by the Th\" uringer Landesstern- warte Tautenburg.}

\InterestConflict{The authors declare that they have no conflict of interest.}


\end{multicols}
\end{document}